\documentclass[10pt,amsmath,amssymb,twocolumn,prb,superscriptaddress,notitlepage]{revtex4-2}

\usepackage{soul}
\usepackage{floatrow}
\usepackage{float}   
\usepackage{subfig}
\usepackage{ragged2e}
\usepackage{braket}
\usepackage{graphicx}
\usepackage[percent]{overpic}
\usepackage{dcolumn}
\usepackage{comment}
\usepackage{xcolor}
\usepackage{color}
\usepackage[colorlinks=true, linkcolor=blue]{hyperref}
\newcommand\Tr{\mathrm{Tr}}

\usepackage[normalem]{ulem}
\newcommand\redsout{\bgroup\markoverwith{\textcolor{red}{\rule[0.5ex]{2pt}{0.4pt}}}\ULon}

\newcommand{\comments}[1]{}

\newcommand{\g}{\vec{\mathbf{g}}}
\newcommand{\V}{\hat{V}}
\newcommand{\U}{\hat{U}}

\DeclareCaptionFormat{justified}{\justifying #1#2#3\par}
\begin{document}

\title{Informationally Complete Distributed Metrology Without a Shared Reference Frame}

\author{Hua-Qing Xu}
\thanks{These two authors contributed equally}
\affiliation{Laboratory of Quantum Information, University of Science and Technology of China, Hefei 230026, China.}
\affiliation{Anhui Province Key Laboratory of Quantum Network, Hefei, Anhui 230026, China.}
\affiliation{CAS Center For Excellence in Quantum Information and Quantum Physics, University of Science and Technology of China, Hefei, Anhui 230026, China.}
\affiliation{Hefei National Laboratory, Hefei 230088, China}
\author{Gong-Chu Li}
\thanks{These two authors contributed equally}
\affiliation{Laboratory of Quantum Information, University of Science and Technology of China, Hefei 230026, China.}
\affiliation{Anhui Province Key Laboratory of Quantum Network, Hefei, Anhui 230026, China.}
\affiliation{CAS Center For Excellence in Quantum Information and Quantum Physics, University of Science and Technology of China, Hefei, Anhui 230026, China.}
\affiliation{Hefei National Laboratory, Hefei 230088, China}
\author{Xu-Song Hong}
\affiliation{Laboratory of Quantum Information, University of Science and Technology of China, Hefei 230026, China.}
\affiliation{Anhui Province Key Laboratory of Quantum Network, Hefei, Anhui 230026, China.}
\affiliation{CAS Center For Excellence in Quantum Information and Quantum Physics, University of Science and Technology of China, Hefei, Anhui 230026, China.}
\author{Lei Chen}
\author{Si-Qi Zhang}
\affiliation{Laboratory of Quantum Information, University of Science and Technology of China, Hefei 230026, China.}
\affiliation{Anhui Province Key Laboratory of Quantum Network, Hefei, Anhui 230026, China.}
\affiliation{CAS Center For Excellence in Quantum Information and Quantum Physics, University of Science and Technology of China, Hefei, Anhui 230026, China.}
\author{Yuancheng Liu}
\affiliation{Laboratory of Quantum Information, University of Science and Technology of China, Hefei 230026, China.}
\affiliation{Anhui Province Key Laboratory of Quantum Network, Hefei, Anhui 230026, China.}

\author{Geng Chen}
\email{chengeng@ustc.edu.cn}
\affiliation{Laboratory of Quantum Information, University of Science and Technology of China, Hefei 230026, China.}
\affiliation{Anhui Province Key Laboratory of Quantum Network, Hefei, Anhui 230026, China.}
\affiliation{CAS Center For Excellence in Quantum Information and Quantum Physics, University of Science and Technology of China, Hefei, Anhui 230026, China.}
\affiliation{Hefei National Laboratory, Hefei 230088, China}

\author{Chuan-Feng Li}
\affiliation{Laboratory of Quantum Information, University of Science and Technology of China, Hefei 230026, China.}
\affiliation{Anhui Province Key Laboratory of Quantum Network, Hefei, Anhui 230026, China.}
\affiliation{CAS Center For Excellence in Quantum Information and Quantum Physics, University of Science and Technology of China, Hefei, Anhui 230026, China.}
\affiliation{Hefei National Laboratory, Hefei 230088, China}

\author{Guang-Can Guo}
\affiliation{Laboratory of Quantum Information, University of Science and Technology of China, Hefei 230026, China.}
\affiliation{Anhui Province Key Laboratory of Quantum Network, Hefei, Anhui 230026, China.}
\affiliation{CAS Center For Excellence in Quantum Information and Quantum Physics, University of Science and Technology of China, Hefei, Anhui 230026, China.}
\affiliation{Hefei National Laboratory, Hefei 230088, China}

\begin{abstract}
In quantum information processing, implementing arbitrary preparations and measurements on qubits necessitates precise information to identify a specific reference frame (RF). In space quantum communication and sensing, where a shared RF is absent, the interplay between locality and symmetry imposes fundamental restrictions on physical systems. 
A restriction on realizable unitary operations results in a no-go theorem prohibiting the extraction of locally encoded information in RF-independent distributed metrology.
Here, we propose a reversed-encoding method applied to two copies of local-unitary-invariant network states. This approach circumvents the no-go theorem while simultaneously mitigating decoherence-like noise caused by RF misalignment, thereby enabling the complete recovery of the quantum Fisher information (QFI). Furthermore, we confirm local Bell-state measurements as an optimal strategy to saturate the QFI. Our findings pave the way for the field application of distributed quantum sensing, which is inherently subject to unknown RF misalignment and was previously precluded by the no-go theorem. 

\end{abstract}

\maketitle

\section{Introduction}
In classical information theory, Shannon’s coding theorems are indifferent to means of encoding, rendering the value of the classical bit fungible, which suffices for various information processing tasks such as data compression, key distribution, or integer factoring \cite{Bartlett2007.RevModPhys.79.555}. The quantum analogs of these tasks might appear to require only the fungible information of qubits. However, this is not the case in quantum information processing, as accessing fungible information demands the selection of specific degrees of freedom for implementing arbitrary preparations and measurements. In other words, the identification of the reference frame (RF) serves as the infungible information. For simplicity, most quantum information protocols assume that the involved parties share a common RF. Nevertheless, this assumption often fails in distributed quantum tasks, ranging from space interferometry --- where a network of sensors measures a single global parameter from a distant source \cite{kovalevsky2012reference, quirrenbach2001optical}, to creating a `world clock' by synchronizing sensors that each measure an independent local parameter \cite{Komar2014.NatPhys.10.582}. In such scenarios, distributed quantum networks promise precision scaling with the number of sites ($N$), as $1/N$ (the Heisenberg limit, HL), surpassing the $1/\sqrt{N}$ standard quantum limit (SQL) of independent sensors \cite{Guo2020.NatPhys.16.1, Zhao2021.PhysRevX.11.031009, Liu2021.NatPhoton.15.137, Proctor2018.PhyRevLett.120.080501, Ge2018.PhysRevLett.121.043604, gessner2018sensitivity, yang2024quantum}. The absence of a shared RF, however, is a primary obstacle to achieve this quantum-enhanced precision.

The role of RF in quantum information processing is closely tied to two essential properties of the physical world: locality and symmetry. Their interplay imposes fundamental restrictions on physical systems. A fundamental restriction revealed by Iman Marvian is that \cite{Marvian2022.NatPhys.18.283}: generic symmetric unitaries cannot be implemented, even approximately, using local symmetric unitaries. 
For RF-independent metrology, this restriction manifests as a no-go theorem: if a parameter-encoding process is locally constructible (i.e., realizable by a sequence of local symmetric operations), the encoded information will be completely erased by RF misalignment. This no-go theorem renders the field application of distributed metrology protocols an intractable task, such as satellite-based network sensing.

Another challenge for RF-independent metrology is that the RF misalignment can be modeled as a \textit{superselection rule} or an effective \textit{decoherence} noise \cite{Bartlett2007.RevModPhys.79.555}, which is particularly detrimental to quantum sensing \cite{Giovannetti2004.Science.306.1330, Giovannetti2006.PhysRevLett.96.010401, Paris2009.IntJQuantumInform.07.125, Giovannetti2011.NatPhoton.96.222, Toth2014.JPhysAMathTheor.47.424006, Jonathan2015.JLightwaveTechnol.33.2359, Pezze2018.RevModPhys.90.035005, Braun2018.RevModPhys.90.035006, Pirandola2018.NatPhoton.12.724} and devastate the quantum advantages \cite{Escher2011.NatPhys.7.406, Demkowicz2012.NatCommun.3.1063, len2022quantum}. 
In principle, aligning the RFs could obviate this noise; however, this may not be feasible for time-sensitive quantum tasks and often requires additional resources and communication overhead \cite{Bartlett2007.RevModPhys.79.555}.
Quantum error correction code \cite{arrad2014increasing, kessler2014quantum, dur2014improved, zhou2018achieving} or decoherence-free subspaces can also mitigate this decoherence noise \cite{Bartlett2007.RevModPhys.79.555}, but the experimental complexity impedes the flexibility and universality of these protocols.

A recent study proposed that utilizing multiple copies of a quantum state can effectively mitigate the decoherence-like effect arising from RF misalignment \cite{Imai2024.arXiv.2410.10518}, thereby enabling the estimation of a global parameter in an RF-independent configuration when it is linked to the properties of entangled states. However, this approach remains fundamentally constrained by the limitations of the no-go theorem.
Simply protecting the state is not enough; one must still be able to encode a parameter onto it. This leaves a critical gap: a method is needed that not only protects the state from RF-induced noise but also provides a valid mechanism for local information encoding that circumvents the no-go theorem.

In this work, we develop a framework to address the crucial challenges for distributed metrology when lacking a shared RF, thereby the no-go theorem can be effectively circumvented and the encoded parameters can be accessed with local operations.
Specifically, we introduce a protocol termed 2-LUI-RE, which employs reversed encoding on two copies of a local unitary invariant (LUI) network state. Fisher information analysis reveals that 2-LUI-RE could fully recover the quantum Fisher information (QFI) and preserve Heisenberg-limited scaling, provided the sites share a Greenberger–Horne–Zeilinger (GHZ) state. Furthermore, we show that local Bell-state measurements (LBM) constitute the optimal measurement strategy for saturating the QFI, whereas standard randomized measurements suffer an exponential loss of information.

\begin{figure*}[ht]
    \centering
    \includegraphics[width=0.9\linewidth]{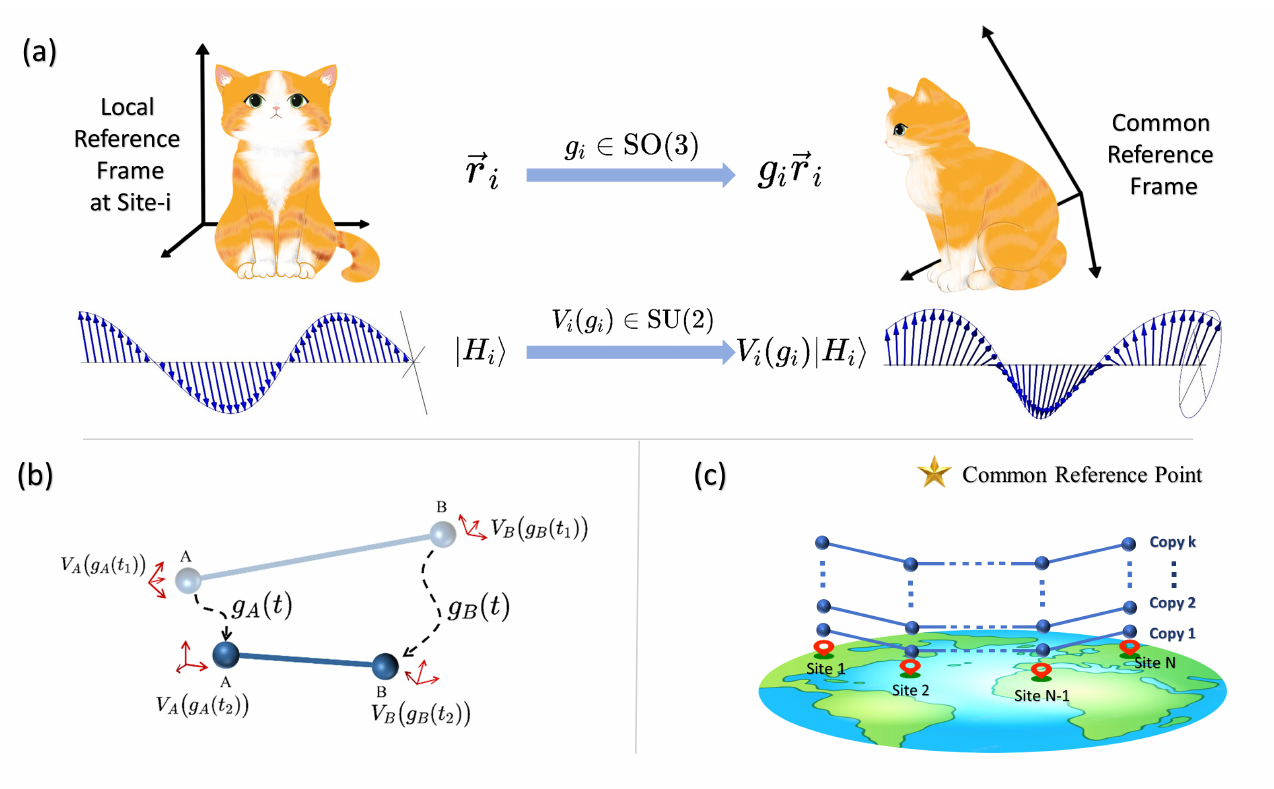}
    \caption{
    \textbf{The Role of RFs and the Impact of Their Misalignment in Quantum Information Processing.} 
    \textbf{(a) Demonstration of RF Misalignment for Classical and Quantum Objects.} The left side shows objects defined in a local RF at Site-$i$, while the right side shows the same objects as viewed from a common RF. For a classical object (upper), e.g., a cat, a misalignment between frames is described by a 3D rotation $g_i \in \mathrm{SO(3)}$. An orientation vector $\vec{\mathbf{r}}_i$ in the local frame is perceived as $g_i\vec{\mathbf{r}}_i$ in the common frame. For a quantum object (down), such as a photon in a horizontally-polarized state $|H_i\rangle$, the same physical rotation corresponds to a transformation $\V_i(g_i) \in \mathrm{SU(2)}$ on its quantum state. This results in a different state, such as an elliptical polarization, in the common frame.
    \textbf{(b) Demonstration of the Effects of Drifting RFs on the State Shared in Two Sites.} From the view of the common reference point, the coordination of site A and B is changing with $g_\text{A}(t),g_\text{B}(t)\in \mathrm{SO(3)}$. The corresponding shared state $\rho_{\text{AB}}$ is experiencing the unitary rotation given $\V_\text{A}(g_\text{A})\otimes \V_\text{B}(g_\text{B})$. The average effect over the measuring time is the $G$-twirling. 
    \textbf{(c) Demonstration of $\boldsymbol{k}$-copy Quantum States Shared in $\boldsymbol{N}$ sites.} The connected dots represent an $N$-qudit state shared in $N$ sites. And the shared $k$ copies mean that $k$ $N$-qudit states are included. At each site, there are $k$ qudits ($k$ balls), one for each copy. }
    \label{fig: demo}
\end{figure*}

\section{Results}

\subsection{The Role of RF in Distributed Quantum Network }
The orientation of any classical object is defined relative to a local RF. When viewed from a different, global reference frame, this local frame appears to be rotated. Any such misalignment between a local and a global frame can be described by an element of the rotation group, SO(3), representing a physical rotation in three-dimensional space (Fig.~\ref{fig: demo}(a)). Similarly, in distributed systems, the mismatch between the local RF at the $i$-th site and a chosen global reference frame is described by a rotation $g_i \in \mathrm{SO}(3)$.

Accurate knowledge of the reference frame alignment is also a prerequisite for quantum information processing tasks. 
Quantum states are invariably defined with respect to the same local RF as the classical apparatus used to prepare and measure them.
For instance, in photonic systems, the RF is typically defined relative to the optical table: horizontal polarization ($|H\rangle$)  refers to polarization parallel to the table surface, and vertical polarization ($|V\rangle$) refers to polarization perpendicular to it, assuming the propagation direction lies along the plane of the table \cite{born2013principles}. 
A qubit state defined by polarizations at the $i$-th site viewed by itself can be written as $|\psi\rangle = \alpha_i|H_i\rangle + \beta_i|V_i\rangle$, where $|H_i\rangle$ and $|V_i\rangle$ denote the local horizontal and vertical polarizations. From a global reference point, this state appears as a different state which has been rotated via a unitary transformation $\V_i(g_i) \in \mathrm{SU}(2)$, such that $|H\rangle = \V_i(g_i)|H_i\rangle$ and $|V\rangle = \V_i(g_i)|V_i\rangle$ \cite{Bartlett2007.RevModPhys.79.555}. This mapping, $\V_i:\mathrm{SO(3)}\rightarrow\mathrm{SU}(d)$ is related to the specific experimental arrangement and provides the intrinsic connection between a physical rotation of the classical apparatus and the corresponding unitary transformation on the quantum state's Hilbert space \cite{Bartlett2007.RevModPhys.79.555}.

\subsection{Decoherence-like Noise by RF Misalignment}
In dynamic scenarios — e.g., satellite-based quantum networks — RFs may fluctuate with time, modeled as $g_i(t)$ (shown in Fig.~\ref{fig: demo}(b)). For a system consist of two sites --- A and B, from a fixed reference point, the effective state over a time interval is given by the averaged state $\rho_{\text{AB},g} = \int dg_\text{A}dg_\text{B} \V_\text{A}(g_\text{A})\otimes \V_\text{B}(g_\text{B})\rho_{\text{AB}}\V_\text{A}(g_\text{A})^\dagger\otimes \V_\text{B}(g_\text{B})^\dagger = \mathcal{G}(\rho_{\text{AB}})$. 
The averaging process, denoted $\mathcal{G}$, is known as a $G$-twirling channel \cite{Bartlett2007.RevModPhys.79.555}. For a general $N$-site state $\rho$, the misalignment is described by a vector of rotations $\g=(g_1,g_2,...,g_N)$, and the twirled state is $\mathcal{G}(\rho)=\int d\g \V(\g)\rho \V(\g)^\dagger$, where $\V(\g) = \V_1(g_1) \otimes \cdots \otimes \V_N(g_N)$. In the extreme case where the local rotations are uniformly distributed over the Haar measure, this $G$-twirling operation becomes a full depolarization channel, mapping any input state to the maximally mixed state, $\mathcal{G}(\rho)=\hat{I}/d^N$, and thereby erasing all initial information.

Fortunately, the effect of RF misalignment can be transformed by sharing multiple copies of a state across sites. When $k$ copies of the state are distributed across $N$ sites, the whole $k$-copy state can be expressed as $\rho^{\otimes k}$ with each $\rho$ an $N$-qudit state (shown in Fig.~\ref{fig: demo}(c)). 
When $k$ copies are shared, we assume that all $k$ qudits at a given site share a common local RF. Consequently, an RF misalignment at site $i$ subjects all $k$ copies to an identical, collective unitary rotation, $\V_i(g_i)^{\otimes k}$. The total effective state, averaged over all misalignments, is given by the $k$-copy $G$-twirling channel, $\mathcal{G}^{(k)}(\rho^{\otimes k})=\int d\g \V(\g)^{\otimes k}\rho^{\otimes k} \V(\g)^{\dagger\otimes k}$. Crucially, for $k\ge 2$, this channel is no longer a simple depolarization. Instead, $\mathcal{G}^{(k)}$ acts as a projection onto the subspace of operators invariant under collective local rotations.

\subsection{No-go Theorem on Distributed Quantum Sensing}
Distributed quantum sensing is inherently associated with two essential properties of the physical world: locality and symmetry, and is therefore subject to the fundamental limitation proposed in Ref.~\cite{Marvian2022.NatPhys.18.283}. Such a limitation defines the implementability of generic symmetric untiaries, stating as: in the continuous system like $\mathrm{SU(2)}$, if a unitary $\U\notin \mathcal{U}_l^G$, then it cannot be implemented using $l$-local symmetric unitaries; on the other hand, if $\U\in \mathcal{U}_l^G$, then it can be implemented with a uniformly finite number of such unitaries \cite{Marvian2022.NatPhys.18.283, d2021introduction}. 
Here, $\mathcal{U}_l^G $ represents the set of all unitary operations obeying a global symmetry that can be generated using interactions acting on at most $l$ sites, which can be expressed as $\mathcal{U}_l^G=\{\U\in l\text{-local unitary }|[\U,\V(\g)]=0,\forall \g \in G\}$, forming a connected and compact Lie group and a closed manifold. Marvian also states that the ``reach" of these operations, measured by the dimension of their manifold (as `dim' below), strictly increases with the $l$-locality, $\dim(\mathcal{U}_l^{G})> \dim (\mathcal{U}_{l'}^G)$ if $l> l'$.

The absence of a shared RF imposes such a symmetry: a physically non-trivial implementable encoding should respect unknown 3D space rotations described by $\mathrm{SO(3)}$. Applying Marvian's restriction to RF-independent scenarios results in a no-go theorem, which states that any information encoded through a 1-local process is completely erased under RF-averaging. Consequently, the accessible information has to be encoded in the set of operations that can only be implemented via non-local approaches, which features $l\ge 2$.

This stringent constraint applies directly when operations are identical across multiple copies, as the overall process remains symmetric. However, the multi-copy framework itself introduces a new physical resource: the ability to apply distinct or correlated operations to each copy. This breaks the SWAP symmetry in copy space, providing a potential pathway to encode information using only local site operations ($l=1$) while circumventing the no-go theorem's core symmetry assumption. More discussions are provided in Supplementary Note I.

~

\subsection{Mitigating RF Decoherence with Multi-copy Twirling}

Based on the above discussion, we conclude that all effects of RFs are generated by local SU($d$) unitaries. And the effect of RF-misalignment acts as a decoherence-like noise described by $G$-twirling.
Thus, for a state to be invariant in the RF-independent case, it must be LUI. 
The group of transformations corresponding to RF misalignment for $k$-copies shared among $N$ sites is therefore $\mathcal{V}_k=\{\V^{\otimes k}|\V\in \mathrm{SU}(d)^{\otimes N}\}$ \cite{Bartlett2007.RevModPhys.79.555, Imai2024.arXiv.2410.10518}. A state $\tilde{\rho}$ is a $k$-copy LUI state if it is a fixed point of this group of operations: $\V_k\tilde{\rho}\V_k^{\dagger}=\tilde{\rho}$ for all $\V_k\in \mathcal{V}_k$.

A general method to construct an LUI state from an arbitrary initial state is to average it over all possible local orientations.  This process, known as $k$-twirling for $k$ copies, projects the state into the RF-invariant subspace, rendering the averaged state immune to subsequent RF noise.  
Specifically, the local $k$-twirling channel at site-$i$ is defined as: $\Phi_i^{(k)}(\cdot)=\int_{\mathrm{Haar}}d\U_i\ \U_i^{\otimes k}(\cdot)\U_i^{\dagger\otimes k}$, where the integral is over the Haar measure of $\mathrm{SU}(d)$  \cite{Elben2022.NatRevPhys.5.9, elben2019statistical}. 

By applying this channel to each site, $\Phi_{\text{local}}^{(k)}=\bigotimes_{i=1}^N \Phi_{i}^{(k)}$, we project the initial $k$-copy state into the LUI subspace, $\tilde{\rho}=\Phi_{\text{local}}^{(k)}(\rho^{\otimes k})$. By construction, the resulting mixed state is invariant under decoherence-noise by any RF misalignment, i.e., $G$-twirling, 
\begin{equation}\label{eq: LUI-unchanged-RF-noise}
    \tilde{\rho} = \mathcal{G}^{(k)}(\tilde{\rho}),\quad \forall\mathcal{G}^{(k)}. 
\end{equation}
For the single-copy case ($k=1$), this projection is trivial, yielding the maximally mixed state $\tilde{\rho}=\hat{I}/d^{N}$. However, for $k\ge 2$, the LUI subspace is non-trivial and can support quantum information.

For quantum metrology, we consider the LUI states are constructed from independent encoded states $\rho_{j,\theta}$ where $j=1,\cdots,k$ indexes the copy. With the encoding process for each copy as $\rho_{j,\theta}=\Theta_{j,\theta}\rho_0\Theta_{j,\theta}^{\dagger}$, the encoded LUI states are
\begin{equation}\label{eq: encoded-LUI-definition}
    \tilde{\rho}_{\theta} = \Phi_{\text{local}}^{(k)}(\rho_{1,\theta}\otimes \cdots \otimes \rho_{k,\theta}).
\end{equation}

The explicit form of the $k$-twirling channel can be derived using Schur-Weyl duality, which decomposes the $k$-fold Hilbert space $\mathcal{H}^{\otimes k}$ into invariant subspaces under the joint action of the unitary and symmetric groups \cite{Zhang2024.arXiv.1408.3782}. For general $O \in \mathbb{C}^{d^k \times d^k}$, the twirling map takes the form $\Phi^{(k)}(O)=\sum_{\sigma,\pi\in S_k}c_{\sigma,\pi}\Tr(OW_{\sigma})W_{\pi}$, where $S_k$ is the permutation group, $W_\sigma$ and $W_\pi$ denote the permutation operators corresponding to $\sigma$ and $\pi$ respectively, and $c_{\sigma,\pi}$ are Weingarten coefficients \cite{Elben2022.NatRevPhys.5.9, elben2019statistical}. 

The two-copy case ($k=2$) is the minimal non-trivial paradigm to preserve the Fisher information in the absence of a shared RF.
For $k = 2$, the resulting state lies in the symmetric and antisymmetric subspaces spanned by the identity and SWAP operators. The SWAP operator $\hat{S}$ acts as $\hat{S} |\psi_1\rangle |\psi_2\rangle = |\psi_2\rangle |\psi_1\rangle$. To be specific, arbitrary matrix $O$ with $\Tr(O)=1$ after the 2-twirling channel satisfies:
\begin{equation}
    \Phi^{(2)}(O) = \frac{1}{d^2-1}\big(\hat{S}+\Tr(\hat{S}O)\big)(\hat{S}-\hat{I}/d).
\end{equation}

Let the 2-copy initial encoded state be $\rho_{1,\theta}$ and $\rho_{2,\theta}$, denoting two identical $N$-particle states parameterized by $\theta$, each defined on a $d$-dimensional local Hilbert space. Define $\mathrm{P}_\theta := \rho_{1,\theta} \otimes \rho_{2,\theta}$. 
The corresponding LUI state is obtained by applying a local 2-twirling channel, which can be implemented locally. The channel factorizes as $\Phi_{\text{local}}^{(2)} = \bigotimes_{i=1}^N \Phi_i^{(2)}$, where each $\Phi_i^{(2)}$ acts locally on the $i$-th site. 
The resulting LUI state becomes 
\begin{equation}\label{Eq: LUI-local}
\begin{aligned}
    \tilde{\rho}_{\theta}&=\Big(\frac{1}{d^2-1}\Big)^N\sum _{\vec{\mathbf{a}}}\Tr(\hat{S}_{\vec{\mathbf{a}}}\mathrm{P}_\theta)\bigotimes_{i:a_i=0}\hat{A}_i\bigotimes_{j:a_j=1}\hat{B}_j,
\end{aligned}
\end{equation}
where $\vec{\mathbf{a}}$ is an $N$-bit binary string, and $\hat{S}_{\vec{\mathbf{a}}} := \bigotimes_{i: a_i=1} \hat{S}_i$ with $\hat{S}_i$ the local SWAP acting on the $i$-th site. The operators $\hat{A}_i := \hat{I}_i - \hat{S}_i/d$ and $\hat{B}_i := \hat{S}_i - \hat{I}_i/d$ define two local subspaces. Further details are provided in Supplementary Note II.

Eq.~\eqref{Eq: LUI-local} shows that the LUI state retains nontrivial dependence on $\theta$ via $\Tr(\hat{S}_{\vec{\mathbf{a}}} \mathrm{P}_\theta)$, in contrast to the trivial case of a single-copy state. Moreover, the structure of the retained information depends on the encoding strategy. 

~

\begin{figure}[t]
    \centering
    \includegraphics[width=0.8\linewidth]{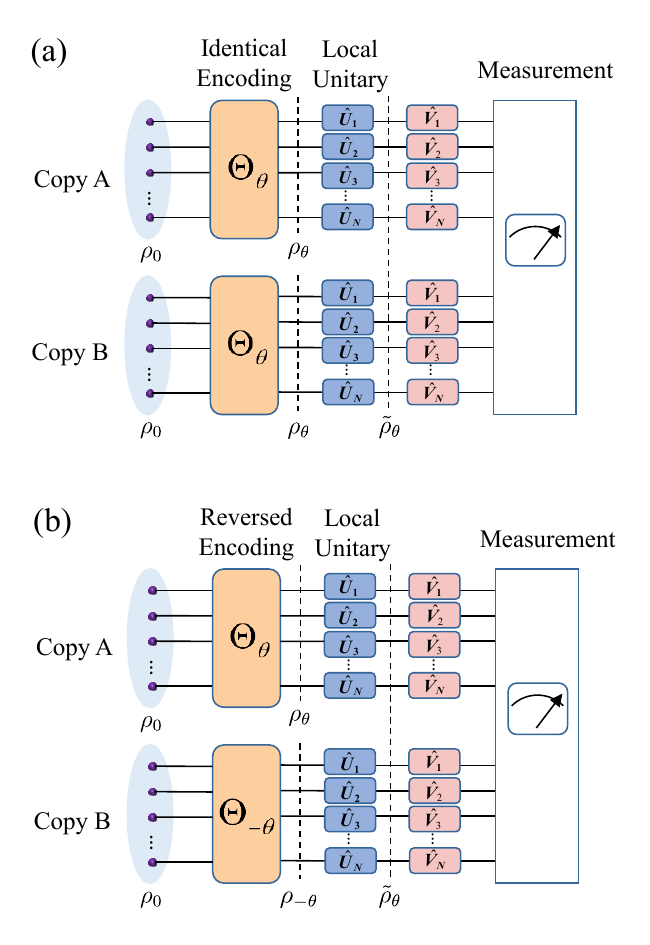}
    \caption{
    \textbf{Comparison of Encoding Strategies for 2-copy LUI States.} \textbf{(a) 2-LUI-IE Protocol:} Both copies undergo the same encoding operation, $\Theta_{\theta}$, and in this case, only nonlocal interactions can give rise to a non-vanishing QFI.
    \textbf{(b) 2-LUI-RE Protocol:} The two copies are reversely encoded with $\Theta_\theta$ and $\Theta_{-\theta}$. This structure explicitly breaks the SWAP symmetry between the copies. Crucially, while $\Theta_{\theta}$ is depicted as a global unitary for generality, this strategy is effective even when the encoding is generated by purely local interactions at a single site.}
    \label{Fig-strategies}
\end{figure}

\subsection{Circumventing the No-go Theorem with 2-LUI-RE Protocol}

In this subsection, we propose a strategy to circumvent the no-go theorem in RF-independent distributed quantum sensing by breaking the SWAP symmetry between the two copies.

For a two-copy LUI state, there are two instinct encoding strategies: (1) Identical Encoding (IE): Following Ref. \cite{Imai2024.arXiv.2410.10518}, both copies are prepared identically: $\mathrm{P}_{\theta}=\rho_{\theta}\otimes \rho_{\theta}$. This construction is manifestly SWAP-symmetric in copy space, i.e. $\Tr(\hat{S}\mathrm{P}_{\theta})=1$. (2) Reversed Encoding (RE): Our proposed strategy prepares the copies with opposite encodings: $\mathrm{P}_{\theta}=\rho_{\theta}\otimes \rho_{-\theta}$. This process explicitly breaks the SWAP symmetry in copy space, i.e., $\Tr(\hat{S}\mathrm{P}_{\theta})$ is related to $\theta$, which, as we will show, is the key to its enhanced metrological power.

To study the metrological advantages of the two strategies, we take rigorous Fisher information analysis. Before proceeding, we briefly review quantum metrology \cite{Paris2009.IntJQuantumInform.07.125}. In quantum parameter estimation, where $\theta$ is encoded in the state $\tilde{\rho}_\theta$, the quantum Cram\'{e}r-Rao bound (QCRB) sets a lower bound on the estimation uncertainty \cite{pang2014quantum}: 
\begin{equation} \label{Eq: QCR}
    \Delta \theta \ge \frac{1}{\sqrt{\nu F}} \ge \frac{1}{\sqrt{\nu \mathcal{F}}} \ge \frac{1}{\sqrt{\nu \mathcal{F}_{\max}}}. 
\end{equation} 
Here, $F$ is the classical Fisher information (CFI), dependent on the measurement $M$, encoding Hamiltonian $\hat{H}$, and input state $\rho_0$. The QFI, $\mathcal{F}$, is the CFI optimized over all measurements, while $\mathcal{F}_{\max}=\max_{\rho_0}\mathcal{F}$ is optimized over probe states and depends solely on the encoding. 

The QFI can be computed from the encoded state $\rho_\theta$ via the symmetric logarithmic derivative (SLD) $\hat{L}_\theta$ satisfying $\partial_{\theta} \rho_\theta = \frac{1}{2}(\rho_\theta \hat{L}_\theta + \hat{L}_\theta \rho_\theta)$, yielding $\mathcal{F} = \Tr(\rho_\theta \hat{L}_\theta^2)$ \cite{Liu_2016}. 
For general mixed states $\tilde{\rho}_\theta = \sum_m \lambda_m |\psi_m\rangle\langle\psi_m|$, the QFI is given by 
\begin{equation}\label{Eq: QFI_definition} 
    \mathcal{F} = \sum_m \frac{(\partial_\theta \lambda_m)^2}{\lambda_m} + 2 \sum_{m,n} \frac{(\lambda_m - \lambda_n)^2}{\lambda_m + \lambda_n} |\langle \partial_\theta \psi_m | \psi_n \rangle|^2, 
\end{equation} 
where the summations exclude terms with $\lambda_m = 0$ or $\lambda_m + \lambda_n = 0$. Since all LUI states are linear combinations of permutation operators (per Eq.~\eqref{Eq: LUI-local}), their eigenstates are $\theta$-independent, and only the first term contributes in Eq.~\eqref{Eq: QFI_definition}.

In the identical encoding strategy, called 2-LUI-IE protocol, as shown in Fig.~\ref{Fig-strategies}(a), the QFI is determined by the eigenvalues of $\tilde{\rho}_\theta$. We denote symmetric or antisymmetric subspaces by a binary vector $\vec{\mathbf{b}}$, and define $\rho_{\theta,\vec{\mathbf{a}}} := \Tr_{i:a_i=0}(\rho_{\theta})$. The QFI reads 
\begin{equation}\label{Eq: identical_encoding_local_unitary} 
    \mathcal{F}_{\text{2-LUI-IE}} = \frac{1}{2^N} \sum_{\vec{\mathbf{b}}} \frac{\left[\sum_{\vec{\mathbf{a}}} (-1)^{\vec{\mathbf{a}}\cdot\vec{\mathbf{b}}} \partial_\theta \Tr(\rho_{\theta,\vec{\mathbf{a}}}^2) \right]^2}{\sum_{\vec{\mathbf{a}}} (-1)^{\vec{\mathbf{a}}\cdot\vec{\mathbf{b}}} \Tr(\rho_{\theta,\vec{\mathbf{a}}}^2)}. 
\end{equation} 
Here, $\Tr(\rho_{\theta,\vec{\mathbf{a}}}^2)$ is related to the 2-Renyi entropy of the subsystem $\hat{B}$ (sites with $a_i=1$). Thus,  as derived in Supplementary Note III, this QFI is sensitive to the entanglement properties of  $\rho_\theta$. Notably, for $\rho_\theta = \Theta_{\theta} \rho_0 \Theta_{\theta}^{\dagger}$, when the encoding operation involves only local interactions -   expressed as $\Theta_{\theta} = \exp(-i\theta \sum_{i} h_i)$ where $h_i$ denotes a local Hamiltonian acting on site $i$ - the 2-Rényi entropy remains independent of $\theta$, and consequently,
\begin{equation}\label{Eq: IE-local-zeroQFI}
    \mathcal{F}_{\text{2-LUI-IE}}=0, \quad \text{if } \Theta_{\theta}\text{ is local.}
\end{equation}
Imai also reaches similar conclusions in \cite{Imai2024.arXiv.2410.10518}, which are consistent with the predictions of the no-go theorem.

In the reversed encoding strategy, called 2-LUI-RE protocol, the two copies are encoded oppositely, as shown in Fig.~\ref{Fig-strategies}(b): $\mathrm{P}_{\theta} = (\Theta_\theta \rho_0 \Theta_\theta^\dagger)\otimes (\Theta_{-\theta} \rho_0 \Theta_{-\theta}^{\dagger})=\rho_{\theta}\otimes \rho_{-\theta}$. For an encoding process $\Theta_{\theta}=e^{-i\hat{H}\theta}$, its reverse is $\Theta_{-\theta}=e^{i\hat{H}\theta}$. 

The reversed operation $\Theta_{-\theta}$ can be realized through several physically relevant methods. In optical systems, such reversal can be achieved by inverting the input-output directions of the encoding black box \cite{chiribella2022quantum}, for instance, by reversing the spatial orientation of the optical crystal. This approach is valid for Hermitian interaction satisfying $\Theta_{-\theta} = \Theta_{\theta}^\dagger$ \cite{guo2024experimental}.
For more general configurations with Hamiltonians expressed in terms of Pauli operators, including most atomic spin systems, a reversal can often be implemented via conjugation with other local unitaries. 
For instance, if $\hat{H}=\frac{1}{2}\sum_{i\in K}Z_i$ on some sites set $K$, then applying a Pauli-X operation $X = \bigotimes_{i=1}^N X_i$ before and after the encoding achieves the reversal: $\Theta_{-\theta} = X\Theta_{\theta}X$.

With the help of the reversed encoding and local-twirling, the QFI of the LUI mixed state $\tilde{\rho}_{\theta}$ is
\begin{equation}\label{Eq: QFI-2-LUI-RE} 
    \mathcal{F}_{\text{2-LUI-RE}} = \frac{1}{2^N} \sum_{\vec{\mathbf{b}}} \frac{\left[\sum_{\vec{\mathbf{a}}} (-1)^{\vec{\mathbf{a}}\cdot\vec{\mathbf{b}}} \partial_\theta \Tr(\rho_{\theta,\vec{\mathbf{a}}} \rho_{-\theta,\vec{\mathbf{a}}}) \right]^2}{\sum_{\vec{\mathbf{a}}} (-1)^{\vec{\mathbf{a}}\cdot\vec{\mathbf{b}}} \Tr(\rho_{\theta,\vec{\mathbf{a}}} \rho_{-\theta,\vec{\mathbf{a}}})}, 
\end{equation} 
where $\rho_{-\theta,\vec{\mathbf{a}}} := \Tr_{i:a_i=0}(\Theta_{-\theta} \rho_0 \Theta_{-\theta}^\dagger)$. 

Moreover, for general cases of $\Theta_{\theta}$, at small $\theta$, the mixed state generated by the 2-LUI-RE protocol approaches the QFI of the untwirled pure state $\mathrm{P}_\theta$, given by $\mathcal{F}_0(\mathrm{P}_\theta) = 8 \left[\Tr(\rho_\theta \hat{H}^2) - \Tr(\rho_\theta \hat{H})^2 \right]$,  leading to the asymptotic equivalence: 
\begin{equation}\label{eq: 2-LUI-RE-QFI} 
\lim_{\theta \to 0} \mathcal{F}_{\text{2-LUI-RE}} = \mathcal{F}_0. 
\end{equation} 
This contrasts starkly with the vanishing QFI in the 2-LUI-IE case, and showcases the optimality of the reversed encoding for small $\theta$. Details of the asymptotic behavior for the 2-LUI-RE protocol are provided in Supplementary Note IV.

Crucially, unlike 2-LUI-IE protocol, $\mathcal{F}_{\text{2-LUI-RE}}$ remains nonzero even when both $\Theta_{\theta}$ and $\Theta_{-\theta}$ are generated by a local Hamiltonian of the form $\hat{H}=\sum_{i\in K}h_i$ with some local Hamiltonian $h_i$ on site $i$, and site set $K$. This is because the overlap terms $\Tr(\rho_{\theta,\vec{\mathbf{a}}} \rho_{-\theta,\vec{\mathbf{a}}})$ are non-trivially dependent on $\theta$. Here, we are going to give three examples with Pauli Hamiltonian. 

\textbf{Example 1: encodings on one site}: $\Theta_{\theta}=e^{-iZ_1\theta/2}$ with $Z_1$ is the local Pauli-Z operator at site-$1$. The QFI is:
\begin{equation}\label{Eq: QFI-onesite}
\begin{aligned}
    \mathcal F&_{\text{2-LUI-RE}}(\rho) 
    =\frac{4\cos^2(\theta)\Tr(\rho^2-Z_1\rho Z_1\rho)}{2-\sin^2(\theta)\Tr(\rho^2-Z_1\rho Z_1\rho)},
\end{aligned}    
\end{equation}
where $\rho$ is an initial encoded pure state. Apparently, for a state $\rho$ with non-trivial initial Fisher information $\mathcal{F}_0=2\Tr(\rho^2-Z_1\rho Z_1\rho)\ne 0$, the QFI after twirling is non-trivial. This case proves that our 2-LUI-RE protocol works for local encodings. And for small $\theta$, $\mathcal F_{\text{2-LUI-RE}}(\rho)$ is near with $\mathcal{F}_0$; for larger $\theta$, the QFI after twirling may decrease. The derivations are provided in Supplementary Note V.

\textbf{Example 2: the initial state is the product state.} Consider the initial product state $|\psi_{\text{prod}}\rangle = \bigotimes_{i=1}^N \frac{1}{\sqrt{2}} (|H_i\rangle+|V_i\rangle)$ defined in the local basis $|H_i\rangle$ and $|V_i\rangle$. When we consider the encoding Hamiltonian $\hat{H}=\frac{1}{2}\sum_{i}Z_i$, the QFI of the resulting LUI state becomes 
\begin{equation}\label{Eq: QFI-product-state}
    \mathcal F_{\text{2-LUI-RE}}(\psi_{\text{prod}})= \frac{4N \cos^2 \theta}{1+\cos^2\theta}.
\end{equation}
In the small-$\theta$ limit, we find $\lim_{\theta\rightarrow0}\mathcal F_{\text{2-LUI-RE}}(\psi_{\text{prod}})=\mathcal F_0(\psi_{\text{prod}})=2N$, which corresponds to the SQL. The derivations are provided in Supplementary Note V. 

\textbf{Example 3: the initial state is the GHZ state.} Now consider the GHZ state in the local basis, $|\psi_{\text{GHZ}}\rangle = \frac{1}{\sqrt{2}}(|H_1H_2\cdots H_N\rangle+|V_1V_2\cdots V_N\rangle)$. Still considering the encoding Hamiltonian  $\hat{H}=\frac{1}{2}\sum_{i}Z_i$, the QFI of the corresponding LUI state is 
\begin{equation}\label{Eq: QFI-GHZ-state}
    \mathcal F_{\text{2-LUI-RE}}(\psi_{\text{GHZ}})= 2N^2 \left[ 1 - \frac{\sin^2 (N\theta)}{\cos^2 (N\theta) + 2^{N-1}} \right].
\end{equation}
Taking the limit $\theta \rightarrow 0$, we obtain
\begin{equation}\label{Eq: QFI-GHZ-state-limit}
    \lim_{\theta\rightarrow0}\mathcal{F}_{\text{2-LUI-RE}}(\psi_{\text{GHZ}})=\mathcal{F}_{0}(\psi_{\text{GHZ}})=2N^2,
\end{equation}
which corresponds to the HL. Importantly, the GHZ state is the optimal probe under local encoding, so the maximal QFI achievable with two independent copies is $\mathcal{F}_{\max} = \mathcal{F}_0(\psi_{\text{GHZ}}) = 2N^2$. Hence, even under local operations and reference-frame averaging, our protocol retains Heisenberg scaling. Further details are provided in Supplementary Note V. 

These examples clearly indicate that the 2-LUI-RE protocol fully preserves the QFI for small values of $\theta$, while for large values of $\theta$, the protocol's sensitivity can be restored to the optimal level through prior knowledge of the system or by implementing an adaptive measurement scheme.

The IE protocol is doubly-constrained by both RF-invariance and SWAP-symmetry, which significantly restrict its manifold of achievable states to the extent that local operations become insufficient. By relaxing the SWAP-symmetry constraint, the RE protocol targets a larger manifold, where the correlated structure of the $\Theta_\theta \otimes \Theta_{-\theta}$ operation provides the necessary resource to encode information robustly in RF-independent scenarios.

In summary, the 2-LUI-RE protocol offers two significant advantages: (i) it enables metrological usefulness of LUI states under fully local operations, and (ii) it provides robustness against reference-frame misalignment while preserving complete QFI. 

\begin{figure}[t]
    \centering
    \includegraphics[width=0.85\linewidth]{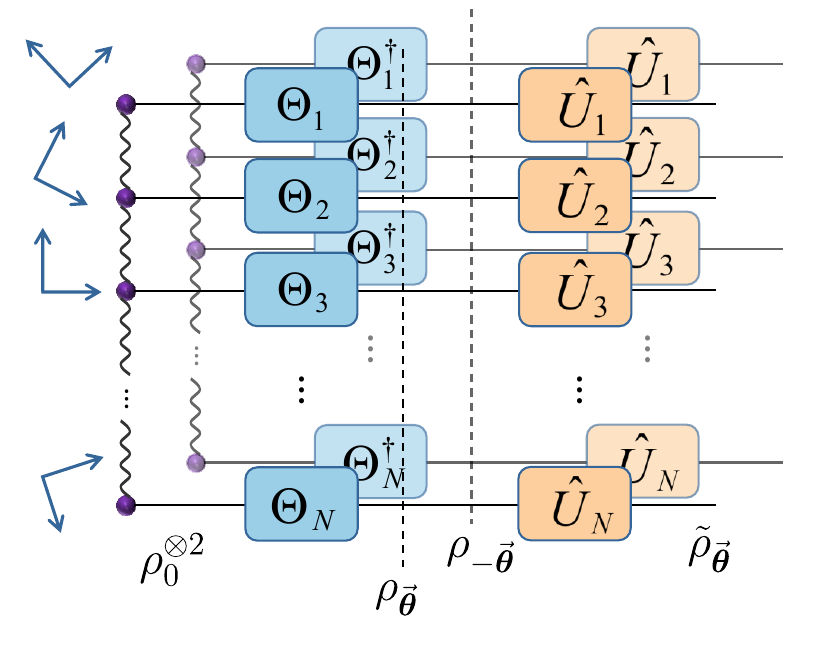}
    \caption{\textbf{2-LUI-RE Protocol with Independent Encodings.} Two copies of the GHZ state ($\rho_0$) are shared by separated sites in a network, and then each site applies reversed-encoding on two on-hand photons with opposing encoding $\Theta_i$ and $\Theta_i^\dagger$, forming two reversely encoded network states $\rho_{\vec{\boldsymbol{\theta}}}$ and $\rho_{-\vec{\boldsymbol{\theta}}}$. Afterward, each site performs local randomized rotations $\U=\bigotimes_{i=1}^N\U_i$ to the two photons to generate the LUI state
    $\tilde{\rho}_{\vec{\boldsymbol{\theta}}}$. The averaged parameter $\bar{\theta}=\frac{1}{N}\sum_{i=1}^N\theta_i$ can be estimated through a specific measurement strategy geared toward $\tilde{\rho}_{\vec{\boldsymbol{\theta}}}$.}
    \label{Fig: distributed-metrology}
\end{figure} 

~

\begin{figure*}[ht]
    \centering
    \includegraphics[width=\linewidth]{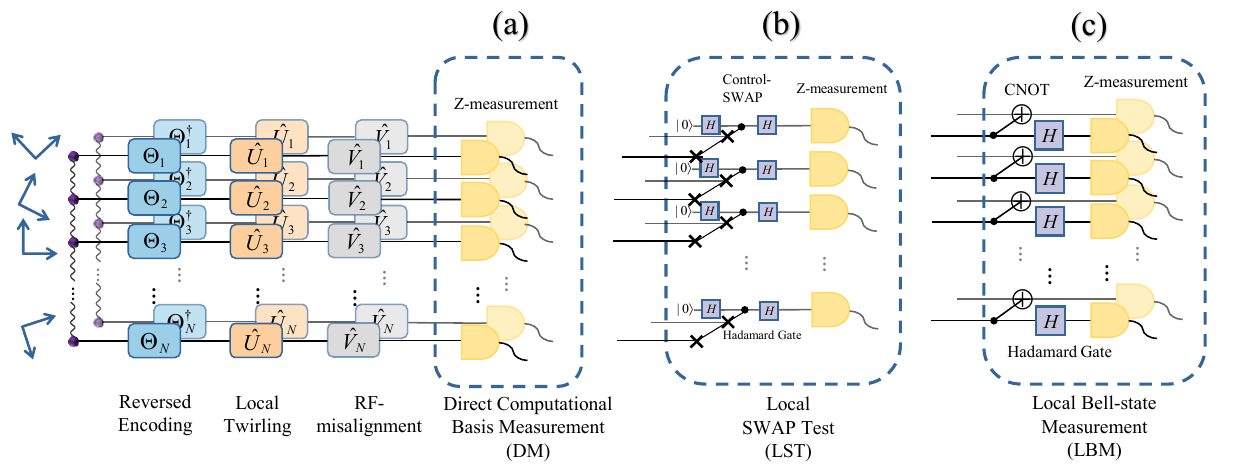}
    \caption{
    \textbf{Local Measurement Strategies for the 2-LUI-RE Protocol.} 
    \textbf{(a) Direct Computational Basis Measurement (DM).} Each qudit is measured directly in the computational basis. This strategy suffers an exponential loss of information.
    \textbf{(b) Local SWAP Test (LST).} An ancillary qubit is introduced at each site to perform a controlled-SWAP operation between the two copies.
    \textbf{(c) Local Bell-state Measurement (LBM).} For qubit systems ($d=2$), the local SWAP test is operationally equivalent to a Bell-state measurement. This is implemented efficiently using a CNOT gate and a Hadamard gate at each site, followed by Z-basis measurements. LST and LBM constitute optimal measurement strategies. }
    \label{fig: measurement}
\end{figure*}

\subsection{Heisenberg-limited Distributed Phase Estimation}

In this subsection, we demonstrate the validity of the 2-LUI-RE protocol in a fundamental task of distributed phase estimation: estimating a single collective parameter that depends on independent local parameters $\theta_i$, each accessed by one of $N$ spatially separated local sensors \cite{Proctor2018.PhyRevLett.120.080501, Ge2018.PhysRevLett.121.043604, gessner2018sensitivity}. A notable paradigm is a `world clock' across a network of clocks across the world \cite{Komar2014.NatPhys.10.582}, which can be modeled as a task to estimate the simple average $\bar{\theta} = \frac{1}{N}\sum_i \theta_i$.

Quantum resources, e.g., the GHZ state, offer a dramatic advantage for this task compared to measuring them individually. Remarkably, a network of $N$ entangled quantum sensors can achieve the HL \cite{Zhao2021.PhysRevX.11.031009, Liu2021.NatPhoton.15.137, Guo2020.NatPhys.16.1, kim2024distributed, malia2022distributed}, where the precision scales as $1/N$ \cite{Ge2018.PhysRevLett.121.043604}. 
Suppose the entire network is in a multi-site GHZ state: $|\psi_{\text{GHZ}}\rangle = \frac{1}{\sqrt{2}} \left( |H_1 H_2 \cdots H_N\rangle + |V_1 V_2 \cdots V_N\rangle \right)$ with $|H_i(V_i)\rangle$ for local horizontal (vertical) polarization. Suppose the parameter of interest at each site is imprinted by a locally-trust interaction with Hamiltonian $h_i = \frac{1}{2}Z_i = \frac{1}{2}\big(|H_i\rangle\langle H_i| - |V_i\rangle\langle V_i|\big)$. The total encoding operation is the product of all local unitaries: $\Theta_{\vec{\boldsymbol{\theta}}} = \exp(-\frac{i}{2} \sum_j Z_j \theta_j)$. When this operation is applied to the GHZ state, the resulting state is:
\begin{equation}
\begin{aligned}
    \Theta_{\vec{\boldsymbol{\theta}}} |\psi_{\text{GHZ}}\rangle 
    &= \frac{1}{\sqrt{2}} \left( |H_1 \cdots H_N\rangle + e^{i N\bar{\theta}} |V_1 \cdots V_N\rangle \right).
\end{aligned}
\end{equation}
The average parameter $\bar{\theta}$ is now encoded as a global phase on the highly sensitive entangled state, which allows it to suit our protocol.

2-LUI-RE protocol uses two copies of the GHZ network state and applies a specific reversed encoding on the second copy before subjecting both to an identical twirling process, as shown in Fig.~\ref{Fig: distributed-metrology}. 
Given the Eq.~\eqref{eq: 2-LUI-RE-QFI}, this protocol preserves the quantum advantage. The QFI for the final, RF-independent state is: 
\begin{equation}
    \lim_{\bar{\theta}\rightarrow 0}\mathcal{F}_{\text{2-LUI-RE}}(\psi_{\text{GHZ}},\bar{\theta}) = \mathcal{F}_{0}(\psi_{\text{GHZ}},\bar{\theta}) = 2N^2,
\end{equation}
which is the original Heisenberg-limited QFI for the ideal GHZ state. This means that even when the sensors have no shared reference frame, our protocol allows them to collectively estimate the global average parameter $\bar{\theta}$ with a precision that scales as $1/N$, the full HL.

~

\begin{figure}[ht]
    \centering
    \includegraphics[width=0.8\linewidth]{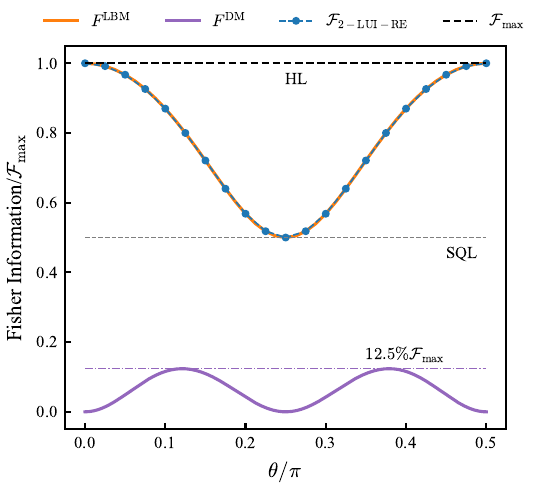}
    \caption{
    \textbf{Fisher Information for Different Measurement Strategies in the 2-LUI-RE Protocol.}  The QFI of the protocol (blue dashed-dotted line) represents the ultimate precision bound. The CFI from the optimal LBM (orange solid line) perfectly saturates the Cram\'{e}r-Rao bound, while the CFI from a naive DM (purple solid line) performs dramatically worse, falling far below the SQL.}
    \label{Fig-CFI-comparision}
\end{figure}

\subsection{Optimal Measurement Strategy}
While the QFI quantifies the ultimate precision bound for a quantum state, this bound is achievable only with optimal measurements. According to Eq.~\eqref{Eq: QCR}, the actual estimation precision after measurement is determined by the CFI, defined via the outcome probabilities $\{p_k\}$ as $F=\sum_{k}(\partial_{\theta} p_k)^2/p_k$. 
For our reversed-encoding strategy, the metrological information is contained in the expectation values of local and global SWAP operators, $\Tr(\hat{S}_{\vec{\mathbf{a}}}\mathrm{P}_{\theta})$. The central challenge is therefore to design a measurement strategy that efficiently extracts these values.

One seemingly straightforward approach is direct computational basis measurement (DM) at each site, as depicted in Fig.~\ref{fig: measurement}(a).
This procedure is operationally equivalent to performing local randomized measurements (LRM) on the initial state $\mathrm{P}_{\theta}$, to estimate the cross-copy correlations $\Tr(\hat{S}_{\vec{\mathbf{a}}}\mathrm{P}_{\theta})$ \cite{elben2020cross, brydges2019probing}.

Although simple to implement, this method is highly inefficient.
Taking the GHZ state with 2-LUI-RE protocol as an example, the resulting CFI is: 
\begin{equation}\label{Eq: CFI-DM}
\begin{aligned}
    F^{\mathrm{DM}} &=\frac{4N^2}{(d+1)^N}\sum_{n=0}^N\binom{N}{n}\frac{\cos^2(N\theta)\sin^2(N\theta)}{d^n+(-1)^n\cos^2(N\theta)}.
\end{aligned}
\end{equation}
As shown in Fig.~\ref{Fig-CFI-comparision}, DM extracts only a small fraction of the maximum QFI $\mathcal{F}_{\max}$. For large $N$, the maximum achievable CFI scales as 
\begin{equation} F^{\mathrm{DM}}_{\max} \sim \frac{1}{d^N} \mathcal{F}_{\max} \sim \frac{N^2}{d^N}, 
\end{equation} demonstrating an exponential loss of information and a failure to preserve the HL. A global randomized measurement performs similarly poorly (demonstrated in Supplementary Note VII).

A far more effective strategy is the local SWAP test (LST) \cite{Buhrman2001.PhysRevLett.87.167902}, which enables explicit extraction of $\langle \hat{S}_{\vec{\mathbf{a}}}\rangle$. As shown in Fig.~\ref{fig: measurement}(b), this is implemented by introducing an ancillary qubit at each site and performing control-SWAP operations.
The power of this method stems from the invariance of the SWAP operator under the twirling channel: $\Tr(\hat{S}_{\vec{\mathbf{a}}}\tilde{\rho}_{\theta}) = \Tr(\hat{S}_{\vec{\mathbf{a}}}\Phi^{(2)}(\mathrm{P}_{\theta})) = \Tr(\Phi^{(2)}(\hat{S}_{\vec{\mathbf{a}}})\mathrm{P}_{\theta}) = \Tr(\hat{S}_{\vec{\mathbf{a}}}\mathrm{P}_{\theta})$. 
Consequently, measuring the ancillary qubits provides direct access to the encoded information.
For measurement basis on $\vec{\mathbf{b}}\in [0,1]^N$, the corresponding measurement probabilities  $p_{\vec{\mathbf{b}}} = \frac{1}{2^N}\sum_{\vec{\mathbf{a}}}(-1)^{\vec{\mathbf{a}}\cdot\vec{\mathbf{b}}}\langle \hat{S}_{\vec{\mathbf{a}}} \rangle$, yielding a CFI that saturates the QFI: 
\begin{equation}
    F^{\mathrm{LST}} = \mathcal{F}_{\text{2-LUI-RE}},\quad \lim_{\theta\rightarrow 0}F^{\mathrm{LST}}=\mathcal{F}_0.
\end{equation}

Especially, for qubit systems ($d=2$), this protocol can be simplified further. Each local SWAP test is operationally equivalent to a local Bell-state measurement, which can be implemented without ancillary systems using CNOT and Hadamard gates (Fig.~\ref{fig: measurement}(c)), which projects the two on-hand photons onto the Bell singlet state, $\frac{1}{\sqrt{2}}(|H_iV_i\rangle - |V_iH_i\rangle)$ and its orthogonal complement.  With this optimal strategy, the gathered CFI matches that of the reversed-encoding LUI state: 
\begin{equation}
    F^{\text{LBM}} = \mathcal{F}_{\text{2-LUI-RE}}, \quad \lim_{\theta\rightarrow 0}F^{\mathrm{LBM}}=\mathcal{F}_0. 
\end{equation}

The stark difference in the efficacy of these measurement strategies is quantified in Fig.~\ref{Fig-CFI-comparision}. The simulation considers an initial $N=2$ GHZ state and an encoding Hamiltonian $\hat{H}=\frac{1}{2}\sum Z_i$. The dashed-dotted blue line shows the QFI of our 2-LUI-RE state ($\mathcal{F}_{\text{2-LUI-RE}}$).
Crucially, the solid orange line ($F^{\text{LBM}}$) shows the CFI extracted using our proposed LBM strategy. It perfectly overlaps with the QFI, proving that LBM is an optimal measurement that saturates the QCRB for all values of $\theta$. When $\theta$ is small, the HL scaling is achieved and for large $\theta$, we could still extract most of the Fisher information with LBM. In stark contrast, the solid purple line ($F^{\text{DM}}$) represents the CFI from a direct computational basis measurement. This simple approach is highly inefficient, peaking at only 12.5\% of the maximum possible information. This performance gap between the optimal LBM and the naive DM becomes larger with the number of particles $N$, underscoring the critical importance of employing the correct measurement strategy to preserve the quantum advantage in distributed sensing. Further details are provided in Supplementary Note VII.

~

\section{Discussion}
Our 2-LUI-RE approach circumvents the no-go theorem on distributed quantum sensing. The reversed encoding breaks the SWAP symmetry between copies and retains the full QFI with local Hamiltonians.  Furthermore, we have identified an optimal measurement strategy --- local SWAP test or local Bell-state measurements --- that can extract this information completely, achieving the HL without requiring shared reference frames or non-local interactions between sites. 

From a practical standpoint, the protocol is experimentally feasible.
Preparing the required 2-copy LUI states is straightforward. For qubit systems, local Clifford operations suffice, and for general $k$-copy states, the twirling can be efficiently approximated using unitary $k$-designs, greatly reducing experimental complexity.

In conclusion, we have presented a comprehensive and experimentally viable solution to the deadlocks of distributed quantum sensing in the absence of a shared reference frame. By simultaneously overcoming the pervasive decoherence effects of RF misalignment and the fundamental limitations on local encoding, our work paves the way for practical implementations of high-precision, networked quantum technologies. 

\section{Data availability}
Authors can confirm that all relevant data are included in the paper and/or its supplementary information files.


%

\section{Acknowledgments} We thank You Zhou for the helpful discussion.
This work was supported by the National Natural Science Foundation of China (Grant Nos.~12350006, 92576202), Quantum Science and Technology-National Science and Technology Major Project (Nos. 2021ZD0301200), and USTC Research Funds of the Double First-Class Initiative (Grant No.~YD2030002026).

\section{Author Contributions} HQ.X. and GC.L. contributed to the framework formulation, the calculation and derivation of key equations, paper writing, and revisions. G.C. contributed to the paper writing and revision.
XS.H., L.C., SQ.Z., Y.L., contributed to the discussion. 
G.C., CF.L., and GG.C. supervised the work.

\section{Competing interests} 
The authors declare no competing interests.

\section{Additional information}
Supplementary Note I-VII

\end{document}


\preprint{}
\title{Supplementary Information: Informationally Complete Distributed Quantum Sensing Without a Shared Reference Frame}

\author{Hua-Qing Xu}
\thanks{These two authors contributed equally}
\affiliation{Laboratory of Quantum Information, University of Science and Technology of China, Hefei 230026, China.}
\affiliation{Anhui Province Key Laboratory of Quantum Network, Hefei, Anhui 230026, China.}
\author{Gong-Chu Li}
\thanks{These two authors contributed equally}
\affiliation{Laboratory of Quantum Information, University of Science and Technology of China, Hefei 230026, China.}
\affiliation{Anhui Province Key Laboratory of Quantum Network, Hefei, Anhui 230026, China.}
\affiliation{CAS Center For Excellence in Quantum Information and Quantum Physics, University of Science and Technology of China, Hefei, Anhui 230026, China.}
\affiliation{Hefei National Laboratory, Hefei 230088, China}
\author{Xu-Song Hong}
\affiliation{Laboratory of Quantum Information, University of Science and Technology of China, Hefei 230026, China.}
\affiliation{Anhui Province Key Laboratory of Quantum Network, Hefei, Anhui 230026, China.}
\affiliation{CAS Center For Excellence in Quantum Information and Quantum Physics, University of Science and Technology of China, Hefei, Anhui 230026, China.}
\author{Lei Chen}
\author{Si-Qi Zhang}
\affiliation{Laboratory of Quantum Information, University of Science and Technology of China, Hefei 230026, China.}
\affiliation{Anhui Province Key Laboratory of Quantum Network, Hefei, Anhui 230026, China.}
\affiliation{CAS Center For Excellence in Quantum Information and Quantum Physics, University of Science and Technology of China, Hefei, Anhui 230026, China.}
\author{Yuancheng Liu}
\affiliation{Laboratory of Quantum Information, University of Science and Technology of China, Hefei 230026, China.}
\affiliation{Anhui Province Key Laboratory of Quantum Network, Hefei, Anhui 230026, China.}

\author{Geng Chen}
\email{chengeng@ustc.edu.cn}
\affiliation{Laboratory of Quantum Information, University of Science and Technology of China, Hefei 230026, China.}
\affiliation{Anhui Province Key Laboratory of Quantum Network, Hefei, Anhui 230026, China.}
\affiliation{CAS Center For Excellence in Quantum Information and Quantum Physics, University of Science and Technology of China, Hefei, Anhui 230026, China.}
\affiliation{Hefei National Laboratory, Hefei 230088, China}

\author{Chuan-Feng Li}
\affiliation{Laboratory of Quantum Information, University of Science and Technology of China, Hefei 230026, China.}
\affiliation{Anhui Province Key Laboratory of Quantum Network, Hefei, Anhui 230026, China.}
\affiliation{CAS Center For Excellence in Quantum Information and Quantum Physics, University of Science and Technology of China, Hefei, Anhui 230026, China.}
\affiliation{Hefei National Laboratory, Hefei 230088, China}

\author{Guang-Can Guo}
\affiliation{Laboratory of Quantum Information, University of Science and Technology of China, Hefei 230026, China.}
\affiliation{Anhui Province Key Laboratory of Quantum Network, Hefei, Anhui 230026, China.}
\affiliation{CAS Center For Excellence in Quantum Information and Quantum Physics, University of Science and Technology of China, Hefei, Anhui 230026, China.}
\affiliation{Hefei National Laboratory, Hefei 230088, China}

\maketitle


\tableofcontents

\newpage
\section{No-go Theorem and Necessary Condition for the Encoding Process}

\subsection{Marian's Theorem in RF-independent Scenario}

The constraints on encoding arise from the interplay between the locality of physical interactions and the symmetries imposed by the problem. We first formalize the hierarchy of symmetric operations.

Marvian's theorem concerns operations constrained by a physical symmetry. In the RF-independent context, the relevant symmetry is the invariance of physical laws under local spatial rotations, described by the group $G=\mathrm{SO(3)}^{\otimes N}$. Let $\mathcal{U}_l^G$ be the Lie group of all unitary operations on an $N$-site Hilbert space ($\mathcal{H}^{\otimes N}$) that can be generated by composing Hamiltonians that are both symmetric with respect to $G$ and at most $l$-local. $\mathcal{U}_l^G = \{ \hat{U} \text{ is } l\text{-local}| [\hat{U}, \hat{V}(\g)]=0,\forall \g\in G \}$ where $\hat{V}(\g)=\bigotimes_{i=1}^N \hat{V}(g_i) \in \mathrm{SU(2)}^{\otimes N}$. 
Here, $\mathcal{U}_l^G$ forms a connected and compact Lie group and a closed manifold.

If a unitary $\hat{U}\notin \mathcal{U}_l^G$, then it cannot be implemented using $l$-local symmetric unitaries; on the other hand, if $\hat{U}\in \mathcal{U}_l^G$, then it can be implemented with a uniformly finite number of such unitaries that is upper bounded by a fixed number that is independent of $\hat{U}$.

Marvian's theorem states a strict inclusion of the corresponding manifolds: 
\begin{equation}\label{eq: Marvian's no-go}
    \dim(\mathcal{U}_{l}^G) > \dim (\mathcal{U}_{l'}^G)\quad  \forall l>l'.
\end{equation}

A more strict analysis based on the Lie algebra, following the Supplementary Note I of \cite{Marvian2022.NatPhys.18.283}. A Lie group $\mathcal{U}$ is characterized by its Lie algebra $\mathfrak{h}$, by $\mathcal{U}=e^{i\mathfrak{h}t}$ with $t\in R$. For the group of symmetric unitaries $\mathcal{U}_l^G$, its real Lie algebra $\mathfrak{h}_l^G$ consists of the corresponding symmetric, Hermitian, $l$-local generators, $\mathfrak{h}_l^G=\{\hat{X} \in l\text{-local}|\hat{X}=\hat{X}^\dagger, [\hat{X},\hat{V}(\g)]=0,\forall g\in G\}$. For $\mathcal{U}^G_l$ is a subgroup of special unitary, $\mathfrak{h}_l^G$ needs to be traceless. 
And $\mathfrak{h}_l^G$ corresponds to the tangent space (at the identity) of the manifold associated to $\mathcal{U}_l^G$.
Thus, the dimension of $\mathcal{U}_l^G$ as a manifold is equal to the dimension of $\mathfrak{h}_l^G$ as a vector space, i.e. $\dim(\mathcal{U}_l^G)=\dim(\mathfrak{h}_l^G)$.
\begin{equation}\label{eq: Lie-algebra-constraint1}
        \dim(\mathfrak{h}_{l}^{G}) > \dim(\mathfrak{h}_{l'}^{ G}),\quad\forall l>l'\\.
\end{equation}

We now extend this formalism to a $k$-copy system on the Hilbert space $(\mathcal{H}^{\otimes N})^{\otimes k}$.
For a $k$-copy system, the space of allowed symmetric operations is larger. Let $\mathcal{U}_{l}^{(k), G}$ be the group of $l$-local, $G$-symmetric operations on the full space. 
Thus, $\mathcal{U}_{l}^{(k), G}=\{\hat{U}_l^{(k)} \text{ is } l\text{-local}|[\hat{U}^{(k)}, \hat{V}(\g)^{\otimes k}]=0,\forall \g\in G\}$.

A physically important subgroup is the set of operations that are also SWAP-symmetric across the copies, which we denote $\mathcal{U}_{l}^{(k), (G,S)}$. That are invariant with arbitrary SWAP operator among copies ---  $\mathcal{U}_{l}^{(k), (G,S)} = \{ \hat{U}^{(k)} \in \mathcal{U}_l^{(k),G} \mid P_\sigma \hat{U}^{(k)} P_\sigma^\dagger = \hat{U}^{(k)}, \forall \sigma \in S_k \}$ with $S_k$ as $k$-permutation group and $P_{\sigma}$ as permutation operator corresponding to $\sigma$. We let $\mathfrak{h}_{l}^{(k), (G,S)}$ as the corresponding Lie algebra. Thus, for every $\hat{H}^{(k)}\in \mathfrak{h}_{l}^{(k), (G,S)}$, it should satisfy $P_{\sigma}\hat{H}^{(k)}P_{\sigma}^\dagger=\hat{H}^{(k)}$. 
Given the Schur-Weyl duality, the generators of the Lie algebra $\mathfrak{h}_l^{(k),(G,S)}$ must be in the form of $\hat{H}^{(k)}=\sum_{j=1}^k\hat{I}^{\otimes j}\otimes \hat{H}\otimes \hat{I}^{k-j}$, where $\hat{H}$ is some Hermitian, $l$-local operator on a single-copy Hilbert space.

By defining $\phi(\hat{H})=\sum_{j=1}^k \hat{I}^{\otimes (j-1)}\otimes \hat{H}\otimes \hat{I}^{\otimes(k-j)}$, we have $[\phi(\hat{H}), \hat{V}(\g)^{\otimes k}]  = \sum_{j=1}^{k} \hat{I}^{\otimes(j-1)} \otimes [\hat{H}, \hat{V}(\g)] \otimes \hat{I}^{\otimes(k-j)}$. Therefore, $[\phi(\hat{H}), \hat{V}(\g)^{\otimes k}]=0$, every term in the sum is zero, $[\hat{H},\hat{V}]=0$, and vice versa. The map $\phi(\hat{H})$ is clearly a linear bijection (an isomorphism) between the two spaces; their dimensions must be equal, i.e.,  
\begin{equation}
    \dim(\mathfrak{h}_{l}^{(k), (G,S)}) = \dim(\mathfrak{h}_{l}^{G}).
\end{equation}

The preceding conclusion allows us to establish a strict hierarchy of operational capabilities. For any $k\ge2$, the set of all RF-invariant operations is strictly larger than its SWAP-symmetric subset, especially with the number of the irreducible representations of the groups, yielding an additional crucial inequality that
\begin{equation}\label{eq: symmetry-manifold-constrain}
\begin{cases}
    \dim(\mathcal{U}_{l}^{(k), G}) > \dim(\mathcal{U}_{l}^{(k), (G,S)}) = \dim((\mathcal{U}_{l}^{G}),\\
    \dim(\mathfrak{h}_{l}^{(k), G}) > \dim(\mathfrak{h}_{l}^{(k), (G,S)}) = \dim(\mathfrak{h}_{l}^{G}).
\end{cases}
\end{equation}

This result demonstrates that relaxing the SWAP-symmetry constraint grants access to a strictly larger manifold of physical operations. Given $\dim(\mathfrak{h}_1^G)=0$, (as 1-local operators cannot be $G$-symmetric in this context), any non-trivial 1-local operation must necessarily violate SWAP-symmetry, i.e., $\hat{U}\in \mathcal{U}_{l\ge 1}^{(k),G}$ but $\hat{U}\notin \mathcal{U}_1^{(k),{(G,S)}}$. Or identical non-local interactions among sites, $\hat{U}\in \mathcal{U}_{l\ge 2}^{(k), (G,S)}$.

\subsection{Necessary Conditions on the Encoding Process}
The symmetric unitaries have a deep connection with the invariant states. And the hierarchy of symmetric operations directly translates into a hierarchy of the invariant state manifolds they can generate.

Let the set of states invariant under a symmetry group $G$ form a manifold $A^G$, $A^G=\{\rho \in \mathcal{H}^{\otimes Nk}|\hat{V}(\g)^{\otimes k}\rho \hat{V}(\g)^{\dagger\otimes k}=\rho, \forall \g\in G\}$. And with additional constraints about SWAP symmetry among copies, the subset $A^{(G,S)}=\{\rho\in A^G| P_{\sigma}\rho P_{\sigma}=\rho, \forall \sigma\in S_k \}$.

If a group of unitary satisfies $[\hat{U},\hat{V}(\g)^{\otimes k}]=0, \forall g\in G$, then, for a $\rho\in A^G$, we have the relation that $\hat{V}(\g)^{\otimes k} \hat{U}\rho \hat{U}^\dagger \hat{V}(\g)^{\dagger\otimes k} = \hat{U}\hat{V}(\g)^{\otimes k} \rho  \hat{V}(\g)^{\dagger\otimes k}\hat{U}^\dagger=\hat{U}\rho \hat{U}^\dagger$. That is to say, any symmetric unitary $\hat{U}\in \mathcal{U}_l^G$ is a symmetry of this manifold, mapping the manifold to itself
\begin{equation}
    \hat{U}A^G\hat{U}^\dagger \subset A^G, \quad \forall \hat{U}\in\mathcal{U}_l^{(k),G}. 
\end{equation}

The connection is made precise by the tangent space. The tangent space of $A^G$ at a point $\rho$, denoted $\mathcal{T}_{\rho}(A^G)$,  characterizes the infinitesimal perturbations that preserve the invariance. For a state $\rho$ to remain invariant under the group action of $\hat{V}(\g)^{\otimes N}$, any perturbation $\hat{X}$ in its tangent space must commute with the group generators. That is to say, for infinitesimal change $\rho\rightarrow \rho+\epsilon \hat{X}$ with small $\epsilon\in R$ and traceless Hermitian $\hat{X}$. $\rho+\epsilon \hat{X} \in A^G$ needs $[\hat{X}, \hat{V}(\g)^{\otimes k}]=0$. Thus, we have $\mathcal{T}_\rho(A^G)=\{\hat{X}|\Tr(\hat{X})=0, \hat{X}=\hat{X}^\dagger, [\hat{X},\hat{V}(\g)^{\otimes k}]=0, \forall g\in G]\}$.
This establishes an isomorphism between the tangent space $\mathcal{T}_{\rho}(A^G)$ and the traceless subspace of the Hermitian algebra $\mathfrak{h}^G$ (that is automatically satisfied for the Lie algebra of the special unitary group). We define it as $\mathcal{T}_{\rho}(A^G)\cong \mathfrak{h}^G$.

We can now define the submanifold of states achievable by $l$-local processes as $A_l^G\subset A^G$. For states in $A_l^G$, $\mathcal{T}_\rho(A_l^G)$ is the set of all traceless operators in $\mathfrak{h_l}^G$. The isomorphism between the tangent space of $A_l^G$ and $\mathcal{U}_l^G$ ensures that the hierarchy of Lie algebras in Eq.~\eqref{eq: Lie-algebra-constraint1} and Eq.~\eqref{eq: symmetry-manifold-constrain} is inherited by the tangent spaces of the invariant state manifolds they generate
\begin{equation}
    \begin{cases}
        \dim(\mathcal{T}_{\rho}(A_{l}^G)) > \dim(\mathcal{T}_{\rho}(A_{l'}^G)),\quad \forall l>l',\\
        \dim(\mathcal{T}_{\rho}(A_{l}^{G})) > \dim(\mathcal{T}_{\rho}(A_{l}^{(G,S)}) = \dim(\mathfrak{h}_l^G).
    \end{cases}
\end{equation}

In this paper, we introduce a construction using $k$-twirling to generate LUI states, i.e., we have $\tilde{\rho}_{\theta}=\Phi^{(k)}_{\text{local}}(\rho_{1,\theta}\otimes \cdots \otimes \rho_{k,\theta}) =\Phi_{\text{local}}^{(k)}(\hat{U}_{\theta}^{(k)}\rho_0^{\otimes k} \hat{U}_{\theta}^{(k)})$, where $\hat{U}_{\theta}^{(k)}$ is copy-independent unitary encoding. We could define the Hamiltonian as $\hat{H}^{(k)}$ with $\hat{U}_{\theta}^{(k)}=\exp({-i\theta \hat{H}^{(k)}})$. For a local $\rho_0$, since $\Phi^{(k)}$ consists of local unitaries, i.e., it is 1-local, if $\hat{H}^{(k)}$ is $l$-local, then we have $\partial_{\theta}(\tilde{\rho}_{\theta})\in \mathcal{T}_{\rho}(A_l^G)$. 

\begin{figure}[t]
    \centering
    \includegraphics[width=0.8\linewidth]{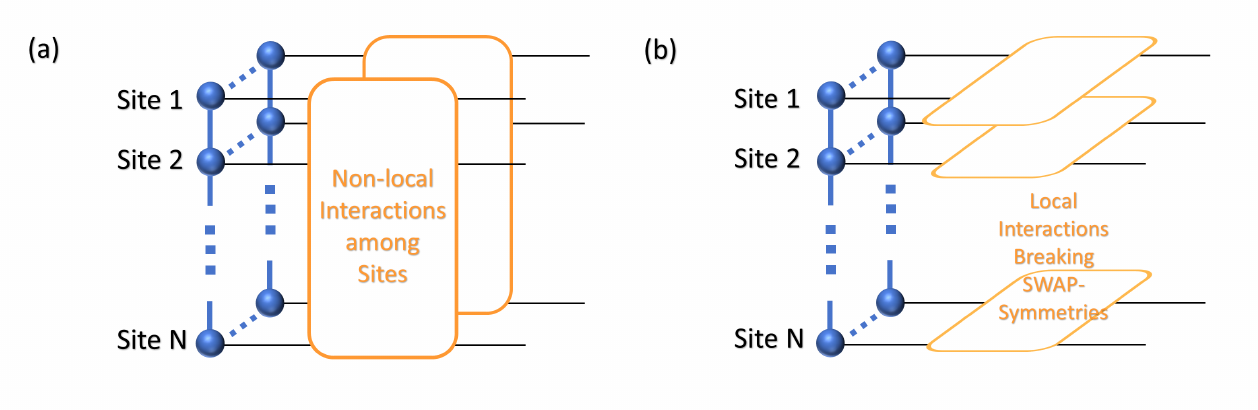}
    \caption{
    \textbf{Demonstration of Constraints on the Non-trivial Interactions.} For an arbitrary interaction, it respects the symmetry of RF rotations. The figure shows two instinctive ways of introducing non-trivial interactions. \textbf{(a) Non-local Interactions among Sites.} The interactions are identical for different copies, but each interaction per copy is a many-body interaction among sites. 
    \textbf{(b) Local Interactions Breaking the SWAP-symmetry among Copies.} The encoding processes are independent among sites, but in each site, the encoding processes break the SWAP symmetry among copies. 
    }
    \label{Fig-constraint}
\end{figure}

Thus, when the encoding processes are identical among copies, the whole encoded states are additionally constrained by the SWAP symmetry. The encoding processes, as a consequence, should still obey the same constraints, i.e., non-local encodings are needed, like in Fig.~\ref{Fig-constraint}(a). Or, multi-copy scenarios provide another possibility --- breaking the SWAP symmetry, which is possible to circumvent the no-go theorem with merely local interactions, like in Fig.~\ref{Fig-constraint}(b). 

Especially, in this paper, we discuss the 2-copy protocol with identical encoding (IE) and reversed encoding (RE) strategies. The IE strategy corresponds to the first case. Its generator $\hat{H}^{(2)}=\hat{H}\otimes \hat{I}+\hat{I}\otimes \hat{H}$ is SWAP-symmetric, with $\hat{H}$ being the Hamiltonian on one copy, confining the velocity vector to the smaller tangent space $\mathcal{T}_{\rho}(A_l^{(G,S)})$. 
A rigorous analysis shows that the information encoded by this protocol is proportional to the change in the local purities of the single-copy state, i.e., terms like $\Tr(\tilde{\rho}_{\theta,\vec{\mathbf{a}}}^2)$. If the single-copy Hamiltonian $\hat{H}$ is 1-local, the local purities are invariant under the evolution, regardless of the initial state's entanglement. Consequently, $\partial_{\theta}\tilde{\rho}_{\theta}=0$. To generate a non-zero velocity on this constrained manifold, $\hat{H}$ must be a many-body interaction ($l \ge 2$) capable of dynamically altering the entanglement between sites.

The RE strategy corresponds to the second case. Its generator $\hat{H}^{(2)}=\hat{H}\otimes \hat{I}-\hat{I}\otimes \hat{H}$ is SWAP-antisymmetric. The process is not confined to the SWAP-symmetric algebra, and its velocity can explore the larger tangent space $\mathcal{T}_{\rho}(A^{(G)})$. As shown by our Fisher Information analysis, the encoded information is related to inter-copy overlap terms like $\Tr(\rho_{\theta,\vec{\mathbf{a}}}\rho_{-\theta,\vec{\mathbf{a}}})$. This quantity is sensitive to $\theta$ even when $\hat{H}$ is 1-local. By breaking the SWAP symmetry, the process accesses a larger manifold of achievable states, where a 1-local $\hat{H}$ is powerful enough to trace a non-trivial path.

\section{Derivation of the 2-copy Local Unitary Invariant State}

In this section, our ultimate goal is to derive the 2-copy local unitary invariant (LUI) state. For the $i$-th particle across both copies, the randomized process is described by the Haar integration $\Phi_i^{(2)}(\mathrm{P}_{\theta})=\int_{\mathrm{Haar}}d\hat{U}_i\;\hat{U}_i^{\otimes 2}\mathrm{P}_{\theta}\hat{U}_i^{\dagger\otimes 2}$ where $\hat{U}_i$ is a unitary rotation acting on the $i$-th particle and $\mathrm{P}_{\theta}$ is defined in the main text. Because of the Schur-Weyl Duality \cite{Zhang2024.arXiv.1408.3782}, this Haar integration has the following result that
\begin{equation}\label{eq:Schur-Weyl one-site}
\begin{aligned}
    \Phi_i^{(2)}(\mathrm{P}_{\theta})=\frac{1}{d^2-1}\left[\operatorname{Tr}_i(\mathrm{P}_{\theta})\otimes\hat{A}_i+\operatorname{Tr}_i(\hat{S}_i\mathrm{P}_{\theta})\otimes\hat{B}_i\right],
\end{aligned}
\end{equation}
where $\hat{A}_i=\hat{I}_i-\hat{S}_i/d$ and $\hat{B}_i=\hat{S}_i-\hat{I}_i/d$ with $\hat{I}_i$ and $\hat{S}_i$ the identity and SWAP operator acting on the two copies of the $i$-th particle. For $N$ particles, the full local unitary process generalize to 
\begin{equation}\label{eq:Schur-Weyl all sites}
\begin{aligned}
    \tilde{\rho}_{\theta}=\Big(\bigotimes_{i=1}^N\Phi_i^{(2)}\Big)(\mathrm{P}_\theta)=\Big(\frac{1}{d^2-1}\Big)^N\sum _{\vec{\mathbf{a}}}\operatorname{Tr}(\hat{S}_{\vec{\mathbf{a}}}\mathrm{P}_\theta)\bigotimes_{i:a_i=0}\hat{A}_i\bigotimes_{j:a_j=1}\hat{B}_j,
\end{aligned}
\end{equation}
where $\vec{\mathbf{a}}$ is defined as an $N$-bit binary string determining whether the $i$-th particle state is $\hat{A}_i$ ($a_i=0$) or $\hat{B}_i$ ($a_i=1$) and $\hat{S}_{\vec{\mathbf{a}}}=\bigotimes_{i:a_i=1}\hat{S}_i$ swaps the two copies of the $i$-th particle if $a_i=1$. This is the general structure of the LUI state $\tilde{\rho}_{\theta}$ under local unitaries. To demonstrate the structure intuitively, we take the case of $N=2$ as an example, where the LUI state is
\begin{equation}\label{eq:LUI for N=2}
\begin{aligned}
    \tilde{\rho}_{\theta}=\Big(\bigotimes_{i=1}^2\Phi_i^{(2)}\Big)(\mathrm{P}_{\theta})=\Big(\frac{1}{d^2-1}\Big)^2\big[\hat{A}_1\otimes\hat{A}_2+\operatorname{Tr}(\hat{S}_1\mathrm{P}_{\theta})\hat{B}_1\otimes\hat{A}_2+\operatorname{Tr}(\hat{S}_2\mathrm{P}_{\theta})\hat{A}_1\otimes\hat{B}_2+\operatorname{Tr}(\hat{S}\mathrm{P}_{\theta})\hat{B}_1\otimes\hat{B}_2\big],
\end{aligned}
\end{equation}
where $\hat{S}$ is the global SWAP operator.

\section{Calculation of the Quantum Fisher Information for LUI States}

In this section, we will calculate the quantum Fisher information (QFI) of the LUI state under identical and reversed encoding processes respectively. According to the method for calculating the QFI of a mixed state \cite{Paris2009.IntJQuantumInform.07.125}, we need to obtain the eigenvalues of the LUI state first, and then calculate the QFI using Eq.~(7) of the main text. 

We first consider the 2-LUI-IE protocol defined in the main text. In this case, we need to find the eigenvalues of Eq.~(5) in the main text. For each eigenvalue, it is the product of the eigenvalues of the $N$ reduced states. From the definition of vectors $\vec{\mathbf{a}}$ and $\vec{\mathbf{b}}$ in the main text, the eigenvalues of the $i$-th particle state have the properties in Tab.~\ref{tab:S1}. We know that if the $i$-th particle state meets $a_i=1$ and $b_i=1$, we will get a coefficient of -1 in the eigenvalue and the total number of -1 is $\vec{\mathbf{a}}\cdot\vec{\mathbf{b}}$. Therefore, the eigenvalues are
\begin{equation}\label{eq:eigenvalue of LUI state}
\begin{aligned}
    \lambda_{\vec{\mathbf{b}}}=\Big(\frac{1}{d^2-1}\Big)^N\Big(\frac{d-1}{d}\Big)^{N-|\vec{\mathbf{b}}|^2}\Big(\frac{d+1}{d}\Big)^{|\vec{\mathbf{b}}|^2}\Big[\sum_{\vec{\mathbf{a}}}(-1)^{\vec{\mathbf{a}}\cdot\vec{\mathbf{b}}}\operatorname{Tr}(\hat{S}_{\vec{\mathbf{a}}}\mathrm{P}_{\theta})\Big],
\end{aligned}
\end{equation}
with a degeneracy of $s_{\vec{\mathbf{b}}}=\Big[\frac{d(d+1)}{2}\Big]^{N-|\vec{\mathbf{b}}|^2}\Big[\frac{d(d-1)}{2}\Big]^{|\vec{\mathbf{b}}|^2}$. As the eigenvalues are independent on $\theta$, the second summation term in Eq.~(7) of the main text vanishes and the QFI can be calculated as
\begin{equation}\label{eq:QFI for 2-LUI}
\begin{aligned}
    \mathcal F_{\mathrm{2-LUI}}=\sum_{\vec{\mathbf{b}}} s_{\vec{\mathbf{b}}}\frac{(\partial_{\theta}\lambda_{\vec{\mathbf{b}}})^2}{\lambda_{\vec{\mathbf{b}}}}=\frac{1}{2^N} \sum_{\vec{\mathbf{b}}} \frac{\Big[\sum_{\vec{\mathbf{a}}}(-1)^{\vec{\mathbf{a}}\cdot\vec{\mathbf{b}}}\partial_{\theta}\operatorname{Tr}(\hat{S}_{\vec{\mathbf{a}}}\mathrm{P}_{\theta})\Big]^2}{\sum_{\vec{\mathbf{a}}}(-1)^{\vec{\mathbf{a}}\cdot\vec{\mathbf{b}}}\operatorname{Tr}(\hat{S}_{\vec{\mathbf{a}}}\mathrm{P}_{\theta})}.
\end{aligned}
\end{equation}
For the 2-LUI-IE protocol, $\operatorname{Tr}(\hat{S}_{\vec{\mathbf{a}}}\mathrm{P}_{\theta})$ reduces to the purity $\operatorname{Tr}(\rho_{\theta,\vec{\mathbf{a}}}^2)$, leading to the QFI expression given in Eq.~(8) of the main text. Notably, this vanishes for local Hamiltonians due to the no-go theorem \cite{Imai2024.arXiv.2410.10518, Marvian2022.NatPhys.18.283}. A brief proof of this is as follows.

We rewrite $\operatorname{Tr}(\rho_{\theta,\vec{\mathbf{a}}}^2)$ in the form of Hamiltonian that
\begin{equation}\label{eq:purity}
    \operatorname{Tr}(\rho_{\theta,\vec{\mathbf{a}}}^2)=\langle\psi_\theta^\text{B}|\langle\psi_\theta^\text{A}|\hat{S}_{\vec{\mathbf{a}}}|\psi_\theta^\text{A}\rangle|\psi_\theta^\text{B}\rangle=\langle\psi_0^\text{B}|\langle\psi_0^\text{A}|e^{i\hat{H}_\text{B}\theta}e^{i\hat{H}_\text{A}\theta}\hat{S}_{\vec{\mathbf{a}}}e^{-i\hat{H}_\text{A}\theta}e^{-i\hat{H}_\text{B}\theta}|\psi_0^\text{A}\rangle|\psi_0^\text{B}\rangle,
\end{equation}
where A, B represent the two copies respectively and $\hat{H}_\text{A}=\hat{H}_\text{B}=\hat{H}$. Then, we have
\begin{equation}\label{eq:partial purity}
    \partial_{\theta}\operatorname{Tr}(\rho_{\theta,\vec{\mathbf{a}}}^2)=i\langle \hat{H}_\text{B}\hat{S}_{\vec{\mathbf{a}}}\rangle+i\langle \hat{H}_\text{A}\hat{S}_{\vec{\mathbf{a}}}\rangle-i\langle\hat{S}_{\vec{\mathbf{a}}}\hat{H}_\text{A}\rangle-i\langle\hat{S}_{\vec{\mathbf{a}}}\hat{H}_\text{B}\rangle,
\end{equation}
where $\langle\cdot\rangle=\langle\psi_\theta^\text{B}|\langle\psi_\theta^\text{A}|\cdot|\psi_\theta^\text{A}\rangle|\psi_\theta^\text{B}\rangle$. If we take a local Hamiltonian that $\hat{H}=\sum_{i=1}^N \hat{H}_i$, the parameter derivative of $\operatorname{Tr}(\rho_{\theta,\vec{\mathbf{a}}}^2)$ vanishes due to $\langle \hat{H}_\text{A}\hat{S}_{\vec{\mathbf{a}}} \rangle=\langle\hat{S}_{\vec{\mathbf{a}}}\hat{H}_\text{A}\rangle$ and $\langle \hat{H}_\text{B}\hat{S}_{\vec{\mathbf{a}}}\rangle=\langle\hat{S}_{\vec{\mathbf{a}}}\hat{H}_\text{B}\rangle$. Then we derive that the term
\begin{equation}\label{eq:zero QFI}
    \frac{\Big[\sum_{\vec{\mathbf{a}}}(-1)^{\vec{\mathbf{a}}\cdot\vec{\mathbf{b}}}\partial_{\theta}\operatorname{Tr}(\rho_{\theta,\vec{\mathbf{a}}}^2)\Big]^2}{\sum_{\vec{\mathbf{a}}}(-1)^{\vec{\mathbf{a}}\cdot\vec{\mathbf{b}}}\operatorname{Tr}(\rho_{\theta,\vec{\mathbf{a}}}^2)}=0,
\end{equation}
whether $\sum_{\vec{\mathbf{a}}}(-1)^{\vec{\mathbf{a}}\cdot\vec{\mathbf{b}}}\operatorname{Tr}(\rho_{\theta,\vec{\mathbf{a}}}^2)$ is zero or not. For the case of $\sum_{\vec{\mathbf{a}}}(-1)^{\vec{\mathbf{a}}\cdot\vec{\mathbf{b}}}\operatorname{Tr}(\rho_{\theta,\vec{\mathbf{a}}}^2)=0$, the L'Hospital's rule also gives the result of Eq.~\eqref{eq:zero QFI}. Therefore, we derive the result that $\mathcal F_{\mathrm{2-LUI-IE}}=0$, which meets the no-go theorem.

\begin{table}[tbp]
    \centering
    \caption{\textbf{Eigenvalues and Degeneracy of the $i$-th Particle State for Local Randomized Process}}
    \label{tab:S1}
    \centering
    \begin{tabular}{cccccc}
        \toprule
        $a_i$ & $i$-th particle state & $b_i$ & Symmetry & Eigenvalue & Degeneracy \\
        \midrule
        \multirow{2}{*}{\centering 0} & \multirow{2}{*}{\centering $\hat{A}_i$} & 0 & symmetric & $\frac{d-1}{d}$ & $\frac{d(d+1)}{2}$ \\
        \cmidrule{3-6}
        & & 1 & anti-symmetric & $\frac{d+1}{d}$ & $\frac{d(d-1)}{2}$ \\
        \midrule
        \multirow{2}{*}{\centering 1} & \multirow{2}{*}{\centering $\hat{B}_i$} & 0 & symmetric & $\frac{d-1}{d}$ & $\frac{d(d+1)}{2}$ \\
        \cmidrule{3-6}
        & & 1 & anti-symmetric & $-\frac{d+1}{d}$ & $\frac{d(d-1)}{2}$ \\
        \bottomrule
    \end{tabular}
\end{table}

For reversed encoding with local unitaries (2-LUI-RE), the QFI retains the structure of Eq.~\eqref{eq:QFI for 2-LUI} but replaces $\operatorname{Tr}(\rho_{\theta,\vec{\mathbf{a}}}^2)$ with $\operatorname{Tr}(\rho_{\theta,\vec{\mathbf{a}}}\rho_{-\theta,\vec{\mathbf{a}}})$, as shown in Eq.~(10) of the main text. The critical distinction is that $\operatorname{Tr}(\rho_{\theta,\vec{\mathbf{a}}}\rho_{-\theta,\vec{\mathbf{a}}})$ depends on $\theta$ even for local encodings, circumventing the no-go theorem.

\section{Proof of the Asymptotic Optimality of QFI for the Reversed Encoding Protocol}

In this section, we try to prove that the QFI is asymptotically optimal for the 2-LUI-RE protocol when estimating a sufficiently small parameter. For convenience, we define $u:=\sum_{\vec{\mathbf{a}}}(-1)^{\vec{\mathbf{a}}\cdot\vec{\mathbf{b}}}\operatorname{Tr}(\rho_{\theta,\vec{\mathbf{a}}}\rho_{-\theta,\vec{\mathbf{a}}})$. The term $\operatorname{Tr}(\rho_{\theta,\vec{\mathbf{a}}}\rho_{-\theta,\vec{\mathbf{a}}})$ can be represented as
\begin{equation}\label{eq:Tr(rho_theta,a rho_-theta,a)}
\begin{aligned}
    \operatorname{Tr}(\rho_{\theta,\vec{\mathbf{a}}}\rho_{-\theta,\vec{\mathbf{a}}})={\langle}\psi_{-\theta}^\text{B}|\langle\psi_{\theta}^\text{A}|\hat{S}_{\vec{\mathbf{a}}}|\psi_{\theta}^\text{A}\rangle|\psi_{-\theta}^\text{B}\rangle=\langle e^{-i\hat{H}_\text{B}\theta}e^{i\hat{H}_\text{A}\theta}\hat{S}_{\vec{\mathbf{a}}}e^{-i\hat{H}_\text{A}\theta}e^{i\hat{H}_\text{B}\theta}\rangle,
\end{aligned}
\end{equation}
where $\langle\cdot\rangle=\langle\psi_0^\text{B}|\langle\psi_0^\text{A}|\cdot |\psi_0^\text{A}\rangle |\psi_0^\text{B}\rangle$ and A, B represent the two copies respectively. Then, we have
\begin{equation}\label{eq:limit partial Tr}
\begin{aligned}
    \lim_{\theta\rightarrow0}\partial_{\theta}\operatorname{Tr}(\rho_{\theta,\vec{\mathbf{a}}}\rho_{-\theta,\vec{\mathbf{a}}})=-i\langle \hat{H}_\text{B}\hat{S}_{\vec{\mathbf{a}}}\rangle+i\langle \hat{H}_\text{A}\hat{S}_{\vec{\mathbf{a}}}\rangle-i\langle\hat{S}_{\vec{\mathbf{a}}}\hat{H}_\text{A}\rangle+i\langle\hat{S}_{\vec{\mathbf{a}}}\hat{H}_\text{B}\rangle.
\end{aligned}
\end{equation}
As $\langle \hat{H}_\text{B}\hat{S}_{\vec{\mathbf{a}}}\rangle=\langle \hat{H}_\text{B}\hat{S}\hat{S}_{\vec{\mathbf{a}}}\hat{S}\rangle=\langle \hat{H}_\text{A}\hat{S}_{\vec{\mathbf{a}}}\rangle$ and $\langle\hat{S}_{\vec{\mathbf{a}}}\hat{H}_\text{A}\rangle=\langle\hat{S}_{\vec{\mathbf{a}}}\hat{H}_\text{B}\rangle$, the limit value $\lim_{\theta\rightarrow0}\partial_{\theta}\operatorname{Tr}(\rho_{\theta,\vec{\mathbf{a}}}\rho_{-\theta,\vec{\mathbf{a}}})=0$ and thus $\lim_{\theta\rightarrow0}\partial_{\theta}u=0$. 

We discuss the term $u$ in detail. First we define the vector $\urcorner\vec{\mathbf{a}}:=\vec{\mathbf{1}}-\vec{\mathbf{a}}$ where $\vec{\mathbf{1}}$ means the $N$-bit vector with all elements 1. Then, we have $\hat{S}_{\vec{\mathbf{a}}}|\psi_0^\text{A}\rangle|\psi_0^\text{B}\rangle=\hat{S}_{\vec{\mathbf{a}}}\hat{S}|\psi_0^\text{A}\rangle|\psi_0^\text{B}\rangle=\hat{S}_{\vec{\mathbf{a}}}^2\hat{S}_{\urcorner\vec{\mathbf{a}}}|\psi_0^\text{A}\rangle|\psi_0^\text{B}\rangle=\hat{S}_{\urcorner\vec{\mathbf{a}}}|\psi_0^\text{A}\rangle|\psi_0^\text{B}\rangle$ which means that
\begin{equation}\label{eq:limit Tr}
    \lim_{\theta\rightarrow0}\operatorname{Tr}(\rho_{\theta,\vec{\mathbf{a}}}\rho_{-\theta,\vec{\mathbf{a}}})=\lim_{\theta\rightarrow0}\operatorname{Tr}(\rho_{\theta,\urcorner\vec{\mathbf{a}}}\rho_{-\theta,\urcorner\vec{\mathbf{a}}}).
\end{equation}
If $|\vec{\mathbf{b}}|^2$ is odd, we have $(-1)^{\vec{\mathbf{a}}\cdot\vec{\mathbf{b}}}+(-1)^{\urcorner\vec{\mathbf{a}}\cdot\vec{\mathbf{b}}}=0$. Therefore, we can get the result that
\begin{equation}\label{eq:zero limit}
\begin{aligned}
    \lim_{\theta\rightarrow0}\Big[(-1)^{\vec{\mathbf{a}}\cdot\vec{\mathbf{b}}}\operatorname{Tr}(\rho_{\theta,\vec{\mathbf{a}}}\rho_{-\theta,\vec{\mathbf{a}}})+(-1)^{\urcorner\vec{\mathbf{a}}\cdot\vec{\mathbf{b}}}\operatorname{Tr}(\rho_{\theta,\urcorner\vec{\mathbf{a}}}\rho_{-\theta,\urcorner\vec{\mathbf{a}}})\Big]=0.
\end{aligned}
\end{equation}
For all $\vec{\mathbf{a}}$, we have the result that $\lim_{\theta\rightarrow0}u=0$. However, if $|\vec{\mathbf{b}}|^2$ is even, the above result is not always correct for all $\vec{\mathbf{b}}$.

We know that the limit value $\lim_{\theta\rightarrow0}\operatorname{Tr}(\rho_{\theta,\vec{\mathbf{a}}}\rho_{-\theta,\vec{\mathbf{a}}})=\operatorname{Tr}(\rho_{0,\vec{\mathbf{a}}}^2)\in[\frac{1}{2},1]$ depends on the entanglement of $\rho_0$. So we define an $N$-particle binary vector $\vec{\mathbf{r}}$ as the entanglement indicator where $r_i=0$ if the $i$-th particle is unentangled and $r_i=1$ if the $i$-th particle is entangled. For any vector $\vec{\mathbf{a}}$, we can always find another vector $\vec{\mathbf{a}}'$ which meets either $\vec{\mathbf{a}}\wedge\vec{\mathbf{r}}=\vec{\mathbf{a}}'\wedge\vec{\mathbf{r}}$ or $\vec{\mathbf{a}}\wedge\vec{\mathbf{r}}+\vec{\mathbf{a}}'\wedge\vec{\mathbf{r}}=\vec{\mathbf{r}}$, where $\wedge$ means the logical bit-wise `and' operation. Under these conditions, we have $\operatorname{Tr}(\rho_{0,\vec{\mathbf{a}}}^2)=\operatorname{Tr}(\rho_{0,\vec{\mathbf{a}}'}^2)$.

For a certain vector $\vec{\mathbf{b}}$, if $\vec{\mathbf{b}}\cdot\vec{\mathbf{r}}<|\vec{\mathbf{b}}|^2$, we can always find two vectors $\vec{\mathbf{a}}$ and $\vec{\mathbf{a}}'$ which meet $\vec{\mathbf{a}}\wedge\vec{\mathbf{r}}=\vec{\mathbf{a}}'\wedge\vec{\mathbf{r}}$ and $(-1)^{\vec{\mathbf{a}}\cdot\vec{\mathbf{b}}}+(-1)^{\vec{\mathbf{a}}'\cdot\vec{\mathbf{b}}}=0$. Then the result can be derived that $(-1)^{\vec{\mathbf{a}}\cdot\vec{\mathbf{b}}}\operatorname{Tr}(\rho_{0,\vec{\mathbf{a}}}^2)+(-1)^{\vec{\mathbf{a}}'\cdot\vec{\mathbf{b}}}\operatorname{Tr}(\rho_{0,\vec{\mathbf{a}}'}^2)=0$. Since the numbers of $\vec{\mathbf{a}}$ and $\vec{\mathbf{a}}'$ are the same, we have the limit value $\lim_{\theta\rightarrow0}u=0$. If $\vec{\mathbf{b}}\cdot\vec{\mathbf{r}}=|\vec{\mathbf{b}}|^2$, the limit value doesn't equal 0. Therefore, the following result can be derived that
\begin{equation}\label{eq:limit u1}
    \lim_{\theta\rightarrow0}\frac{(\partial_{\theta}u)^2}{u}=\begin{cases}0,&\vec{\mathbf{b}}\cdot\vec{\mathbf{r}}=|\vec{\mathbf{b}}|^2\;\&\;|\vec{\mathbf{b}}|^2\;\text{is\;even},\\\frac{0}{0},&\text{other\;}\vec{\mathbf{b}}.\end{cases}
\end{equation}
And we need to use L'Hospital's rule to the terms of $\frac{0}{0}$ that
\begin{equation}\label{eq:limit u2}
\begin{aligned}
    \lim_{\theta\rightarrow0}\frac{(\partial_{\theta}u)^2}{u}=\frac{2\partial_{\theta}u\partial_{\theta}^2u}{\partial_{\theta}u}\Big|_{\theta=0}=2\partial_{\theta}^2u\big|_{\theta=0}=2\sum_{\vec{\mathbf{a}}}(-1)^{\vec{\mathbf{a}}\cdot\vec{\mathbf{b}}}\partial_{\theta}^2\operatorname{Tr}(\rho_{\theta,\vec{\mathbf{a}}}\rho_{-\theta,\vec{\mathbf{a}}})\Big|_{\theta=0}.
\end{aligned}
\end{equation}
To rewrite Eq.~\eqref{eq:limit u2} in the form of Hamiltonian, we get this term
\begin{equation}\label{eq:square partial Tr}
\begin{aligned}
    \partial_{\theta}^2\operatorname{Tr}(\rho_{\theta,\vec{\mathbf{a}}}\rho_{-\theta,\vec{\mathbf{a}}})\big|_{\theta=0}=-2\langle \hat{H}_\text{A}^2\hat{S}_{\vec{\mathbf{a}}}\rangle+2\langle \hat{H}_\text{A}\hat{H}_\text{B}\hat{S}_{\vec{\mathbf{a}}}\rangle+2\langle \hat{S}_{\vec{\mathbf{a}}}\hat{H}_\text{A}\hat{H}_\text{B}\rangle-2\langle \hat{S}_{\vec{\mathbf{a}}}\hat{H}_\text{A}^2\rangle-4\langle \hat{H}_\text{A}\hat{S}_{\vec{\mathbf{a}}}\hat{H}_\text{B}\rangle+4\langle \hat{H}_\text{A}\hat{S}_{\vec{\mathbf{a}}}\hat{H}_\text{A}\rangle.
\end{aligned}
\end{equation}
Then the QFI can be rewritten as
\begin{equation}\label{eq:limit QFI1}
\begin{aligned}
    \lim_{\theta\rightarrow0}\mathcal F_{\mathrm{2-LUI-RE}}=\frac{1}{2^{N-1}}\sum_{\vec{\mathbf{a}}}\Big(\sum_{\vec{\mathbf{b}}}-\sum_{\vec{\mathbf{b}}}{'}\Big)(-1)^{\vec{\mathbf{a}}\cdot\vec{\mathbf{b}}}\partial_{\theta}^2\operatorname{Tr}(\rho_{\theta,\vec{\mathbf{a}}}\rho_{-\theta,\vec{\mathbf{a}}})\Big|_{\theta=0},
\end{aligned}
\end{equation}
where $\sum'$ means the sum for all $\vec{\mathbf{b}}$'s meeting the conditions that $\vec{\mathbf{b}}\cdot\vec{\mathbf{r}}=|\vec{\mathbf{b}}|^2$ and $|\vec{\mathbf{b}}|^2$ is even, with a total number of $2^{|\vec{\mathbf{r}}|^2-1}$. For each $\vec{\mathbf{a}}$, we have $\sum_{\vec{\mathbf{b}}}(-1)^{\vec{\mathbf{a}}\cdot\vec{\mathbf{b}}}=0$ and then $\sum_{\vec{\mathbf{a}}}\sum_{\vec{\mathbf{b}}}(-1)^{\vec{\mathbf{a}}\cdot\vec{\mathbf{b}}}\partial_{\theta}^2\operatorname{Tr}(\rho_{\theta,\vec{\mathbf{a}}}\rho_{-\theta,\vec{\mathbf{a}}})\Big|_{\theta=0}=0$.
Therefore, the QFI has the form that
\begin{equation}\label{eq:limit QFI2}
\begin{aligned}
    \lim_{\theta\rightarrow0}\mathcal F_{\mathrm{2-LUI-RE}}=-\frac{1}{2^{N-1}}\sum_{\vec{\mathbf{a}}}\sum_{\vec{\mathbf{b}}}{'}(-1)^{\vec{\mathbf{a}}\cdot\vec{\mathbf{b}}}\partial_{\theta}^2\operatorname{Tr}(\rho_{\theta,\vec{\mathbf{a}}}\rho_{-\theta,\vec{\mathbf{a}}})\Big|_{\theta=0}.
\end{aligned}
\end{equation}
For a certain $\vec{\mathbf{a}}$, if it meets the condition $0<\vec{\mathbf{a}}\cdot\vec{\mathbf{r}}<|\vec{\mathbf{r}}|^2$, the following result can be derived due to symmetry that 
\begin{equation}\label{eq:zero sum}
    \sum_{\vec{\mathbf{b}}}{'}(-1)^{\vec{\mathbf{a}}\cdot\vec{\mathbf{b}}}\partial_{\theta}^2\operatorname{Tr}(\rho_{\theta,\vec{\mathbf{a}}}\rho_{-\theta,\vec{\mathbf{a}}})\Big|_{\theta=0}=0.
\end{equation}
If $\vec{\mathbf{a}}\cdot\vec{\mathbf{r}}=0$ or $|\vec{\mathbf{r}}|^2$, we have $(-1)^{\vec{\mathbf{a}}\cdot\vec{\mathbf{b}}}=1$ and $\hat{S}_{\vec{\mathbf{a}}}|\psi_0^\text{A}\rangle|\psi_0^\text{B}\rangle=|\psi_0^\text{A}\rangle|\psi_0^\text{B}\rangle$. Then the term can be derived as
\begin{equation}\label{eq:-1 square partial Tr}
\begin{aligned}
    (-1)^{\vec{\mathbf{a}}\cdot\vec{\mathbf{b}}}\partial_{\theta}^2\operatorname{Tr}(\rho_{\theta,\vec{\mathbf{a}}}\rho_{-\theta,\vec{\mathbf{a}}})\big|_{\theta=0}=-4\big(\langle \hat{H}^2\rangle-\langle \hat{H}\rangle^2\big)-4\langle \hat{H}_\text{A}\hat{S}_{\vec{\mathbf{a}}}\hat{H}_\text{B}\rangle+4\langle \hat{H}_\text{A}\hat{S}_{\vec{\mathbf{a}}}\hat{H}_\text{A}\rangle.
\end{aligned}
\end{equation}
The number of $\vec{\mathbf{a}}$ meeting Eq.~\eqref{eq:-1 square partial Tr} is $2^{N-|\vec{\mathbf{r}}|^2+1}$. We know that $\langle \hat{H}_\text{A}\hat{S}_{\vec{\mathbf{a}}}\hat{H}_\text{A}\rangle-\langle \hat{H}_\text{A}\hat{S}_{\vec{\mathbf{a}}}\hat{H}_\text{B}\rangle=\langle \hat{H}_\text{A}\hat{S}_{\vec{\mathbf{a}}}\hat{S}\hat{H}_\text{B}\rangle-\langle \hat{H}_\text{A}\hat{S}_{\vec{\mathbf{a}}}\hat{S}\hat{H}_\text{A}\rangle=\langle \hat{H}_\text{A}\hat{S}_{\urcorner\vec{\mathbf{a}}}\hat{H}_\text{B}\rangle-\langle \hat{H}_\text{A}\hat{S}_{\urcorner\vec{\mathbf{a}}}\hat{H}_\text{A}\rangle$. So the summation of the last two terms in Eq.~\eqref{eq:-1 square partial Tr} for all $\vec{\mathbf{a}}$ is 0. Substituting into Eq.~\eqref{eq:limit QFI2}, we derive the limit of the QFI that
\begin{equation}\label{eq:limit QFI for 2-LUI-RE}
\begin{aligned}
    \lim_{\theta\rightarrow0}\mathcal F_{\mathrm{2-LUI-RE}}=\frac{1}{2^{N-1}}\times2^{N-|\vec{\mathbf{r}}|^2+1}\times2^{|\vec{\mathbf{r}}|^2-1}\times4\big(\langle \hat{H}^2\rangle-\langle \hat{H}\rangle^2\big)=8\big(\langle \hat{H}^2\rangle-\langle \hat{H}\rangle^2\big)=\mathcal F_0.
\end{aligned}
\end{equation}

Above all, we have proved that for the 2-LUI-RE protocol, the QFI of the LUI state can be asymptotically optimal when estimating a sufficiently small parameter $\theta$.

\section{Derivation of the QFIs for Three Examples in the Main Text}

\subsection{Derivation of the QFI for Example 1}

In Example 1 of the main text, we consider the case that the parameter is reversed-encoded on just one site to show the metrological power of the 2-LUI-RE protocol. Without loss of generality, we first consider the case that the parameter is encoded on $m$ sites ($m\leq N$). In this case, all the sites in the network can be decomposed into two parts, the encoded part with $m$ sites and the unencoded part with $N-m$ sites. Then the binary string $\vec{\mathbf{a}}$ defined in the main text can also be decomposed into the sum of two strings that $\vec{\mathbf{a}}=\vec{\mathbf{a}}_1\oplus\vec{\mathbf{a}}_2$ with the element $a_i$ of $\vec{\mathbf{a}}$ in the string $\vec{\mathbf{a}}_1$ if the parameter is encoded on the $i$-th site, and in the string $\vec{\mathbf{a}}_2$ if the parameter is not encoded on the $i$-th site. The same decomposition can be made for the binary string $\vec{\mathbf{b}}$ defined in the main text. 

Now we focus on the QFI, in which the trace term $\Tr(\rho_{\theta,\vec{\mathbf{a}}}\rho_{-\theta,\vec{\mathbf{a}}})$ is significant. For a certain site $i$, if it does not encode the parameter, the trace operation on the two copies of the particle on site $i$ before or after the swap operation can not affect the result of the trace term. Thus we have
\begin{equation}\label{eq:any a2}
    \Tr(\rho_{\theta,\vec{\mathbf{a}}}\rho_{-\theta,\vec{\mathbf{a}}})=\Tr(\rho_{\theta,\vec{\mathbf{a}}_1}\rho_{-\theta,\vec{\mathbf{a}}_1}),\;\;\;\;\; \forall\vec{\mathbf{a}}_2,
\end{equation}
with $\vec{\mathbf{a}}=\vec{\mathbf{a}}_1\oplus\vec{\mathbf{a}}_2$ and $\rho_{\theta,\vec{\mathbf{a}}_1}=\Tr_{i:a_i\in\vec{\mathbf{a}}_1,a_i=0}(\rho_{\theta})$, $\rho_{-\theta,\vec{\mathbf{a}}_1}=\Tr_{i:a_i\in\vec{\mathbf{a}}_1,a_i=0}(\rho_{-\theta})$. In this case, only terms with $\vec{\mathbf{b}}_2=\vec{\mathbf{1}}$ in the sum terms of the QFI expression can take non-zero values. Then we have
\begin{equation}\label{eq:non-zero sum}
    \sum_{\vec{\mathbf{a}}}(-1)^{\vec{\mathbf{a}}\cdot\vec{\mathbf{b}}}\Tr(\rho_{\theta,\vec{\mathbf{a}}}\rho_{-\theta,\vec{\mathbf{a}}})=2^{N-m}\sum_{\vec{\mathbf{a}}_1}(-1)^{\vec{\mathbf{a}}_1\cdot\vec{\mathbf{b}}_1}\Tr(\rho_{\theta,\vec{\mathbf{a}}_1}\rho_{-\theta,\vec{\mathbf{a}}_1}).
\end{equation}
Thus, the QFI for this case can be derived as
\begin{equation}\label{eq:QFI for m sites}
    \mathcal{F}_{\text{2-LUI-RE}}^{m\text{-site}}=\frac{1}{2^m} \sum_{\vec{\mathbf{b}}_1} \frac{\left[\sum_{\vec{\mathbf{a}}_1} (-1)^{\vec{\mathbf{a}}_1\cdot\vec{\mathbf{b}}_1} \partial_\theta \Tr(\rho_{\theta,\vec{\mathbf{a}}_1} \rho_{-\theta,\vec{\mathbf{a}}_1}) \right]^2}{\sum_{\vec{\mathbf{a}}_1} (-1)^{\vec{\mathbf{a}}_1\cdot\vec{\mathbf{b}}_1} \Tr(\rho_{\theta,\vec{\mathbf{a}}_1} \rho_{-\theta,\vec{\mathbf{a}}_1})}.
\end{equation}

Then we go back to the case of Example 1, where we take $m=1$. We can derive this QFI as
\begin{equation}\label{eq:QFI for 1 site}
    \mathcal{F}_{\text{2-LUI-RE}}^{\text{1-site}}=\frac{1}{2}\frac{\big[\partial_{\theta}\Tr(\rho_{\theta}\rho_{-\theta})\big]^2}{1+\Tr(\rho_{\theta}\rho_{-\theta})}+\frac{1}{2}\frac{\big[\partial_{\theta}\Tr(\rho_{\theta}\rho_{-\theta})\big]^2}{1-\Tr(\rho_{\theta}\rho_{-\theta})}=\frac{\big[\partial_{\theta}\Tr(\rho_{\theta}\rho_{-\theta})\big]^2}{1-\big[\Tr(\rho_{\theta}\rho_{-\theta})\big]^2}.
\end{equation}
For the case of Example 1, we take the encoding $\Theta_{\theta}=e^{-iZ_1\theta/2}$ with $Z_1$ the local Pauli-Z operator at site-1, and $\rho_{\theta}=\Theta_{\theta}\rho\Theta_{\theta}^{\dagger}$ with $\rho=|\psi\rangle\langle\psi|$ an initial pure state. Under this condition, we have $\Tr(\rho_{\theta}\rho_{-\theta})=\Tr(\rho^2)-\sin^2(\theta)\Tr(\rho^2-\rho Z_1\rho Z_1)$. Thus, the QFI for the case of Example 1 can be derived as
\begin{equation}\label{eq:QFI for Example 1}
    \mathcal{F}_{\text{2-LUI-RE}}(\rho)=\frac{4\cos^2(\theta)\Tr(\rho^2-\rho Z_1\rho Z_1)}{2-\sin^2(\theta)\Tr(\rho^2-\rho Z_1\rho Z_1)}.
\end{equation}
For a small $\theta$, we have
\begin{equation}\label{eq:limit QFI for Example 1}
    \lim_{\theta\rightarrow0}\mathcal{F}_{\text{2-LUI-RE}}(\rho)=2\Tr(\rho^2-\rho Z_1\rho Z_1)=2\big[1-\Tr(\rho Z_1\rho Z_1)\big]=8\big(\langle (Z_1/2)^2\rangle-\langle Z_1/2\rangle^2\big)=
    \mathcal{F}_0.
\end{equation}
From Eq.~(\ref{eq:QFI for Example 1}) and Eq.~(\ref{eq:limit QFI for Example 1}), we see that a non-zero even the maximum QFI can be obtained just by encoding the parameter on a single site, which proves that the 2-LUI-RE protocol works for local encodings.

\subsection{Derivation of the QFI for Example 2}

In Example 2 of the main text, we prepare two copies of the initial product state $|\psi_{\mathrm{prod}}\rangle= \bigotimes_{i=1}^N \frac{1}{\sqrt{2}} (|H_i\rangle+|V_i\rangle)$ where $|H_i\rangle$ and $|V_i\rangle$ are defined in the main text. After encoding the parameter $\theta$, we have $|\psi_{\pm\theta}\rangle=e^{\mp iZ\theta/2}|\psi_{\mathrm{prod}}\rangle=\bigotimes_{i=1}^N \frac{1}{\sqrt{2}} (|H_i\rangle+e^{\pm i\theta}|V_i\rangle)$.

For the 2-LUI-RE protocol, we have $\operatorname{Tr}(\rho_{\theta,\vec{\mathbf{a}}}\rho_{-\theta, \vec{\mathbf{a}}})=\cos^{2|\vec{\mathbf{a}}|^2}\theta$ according to above definitions. For convenience, we define $n=|\vec{\mathbf{a}}|^2$, $k=|\vec{\mathbf{b}}|^2$ and $m=\vec{\mathbf{a}}\cdot\vec{\mathbf{b}}$. Therefore, we can calculate that
\begin{equation}\label{eq:sum cos}
\begin{aligned}
    \sum_{\vec{\mathbf{a}}}(-1)^{\vec{\mathbf{a}}\cdot\vec{\mathbf{b}}}\cos^{2|\vec{\mathbf{a}}|^2}\theta=\sum_{m=0}^{k}\sum_{n=m}^{N-k+m}\binom{k}{m}\binom{N-k}{n-m}(-1)^m\cos^{2n}\theta=(1+\cos^2\theta)^{N-k}\sin^{2k}\theta.
\end{aligned}
\end{equation}
According to Eq.~(10) of the main text, the QFI is
\begin{equation}\label{eq:QFI for 2-LUI-RE product}
\begin{aligned}
    \mathcal F_{\mathrm{2-LUI-RE}}(\psi_{\mathrm{prod}})&=\frac{1}{2^N}\sum_{k=0}^{N}\binom{N}{k}\frac{\Big\{\partial_{\theta}\big[(1+\cos^2\theta)^{N-k}\sin^{2k}\theta\big]\Big\}^2}{(1+\cos^2\theta)^{N-k}\sin^{2k}\theta}= \frac{4N \cos^2 \theta}{1+\cos^2\theta}.
\end{aligned}
\end{equation}
From Eq.~\eqref{eq:QFI for 2-LUI-RE product}, we can derive that the SQL can be recovered as $\theta\rightarrow0$ that
\begin{equation}\label{eq:SQL limit}
    \lim_{\theta\rightarrow0}\mathcal F_{\mathrm{2-LUI-RE}}(\psi_{\mathrm{prod}})=2N.
\end{equation}

\subsection{Derivation of the QFI for Example 3}

In Example 3 of the main text, we prepare two copies of the initial GHZ state $|\psi_{\mathrm{GHZ}}\rangle=\frac{1}{\sqrt{2}}(|H_1H_2\cdots H_N\rangle+|V_1V_2\cdots V_N\rangle)$. After parameter encoding process, we have $|\psi_{\pm\theta}\rangle=\frac{1}{\sqrt{2}}(|H_1H_2\cdots H_N\rangle+e^{\pm iN\theta}|V_1V_2\cdots V_N\rangle)$. For the case of 2-LUI-RE, we have the result that
\begin{equation}\label{eq:different Tr}
    \operatorname{Tr}(\rho_{\theta,\vec{\mathbf{a}}}\rho_{-\theta,\vec{\mathbf{a}}})=\begin{cases}\frac{1}{2},&0<|\vec{\mathbf{a}}|^2< N,\\1,&|\vec{\mathbf{a}}|^2=0,\\\cos^2(N\theta),&|\vec{\mathbf{a}}|^2=N.\end{cases}
\end{equation}
Then we can derive that
\begin{equation}\label{eq:different sum}
\begin{aligned}
    \sum_{\vec{\mathbf{a}}}(-1)^{\vec{\mathbf{a}}\cdot\vec{\mathbf{b}}}\operatorname{Tr}(\rho_{\theta,\vec{\mathbf{a}}}\rho_{-\theta,\vec{\mathbf{a}}})=
\begin{cases}
    2^{N-1}+\cos^2(N\theta),&|\vec{\mathbf{b}}|^2=0,\\
    1-\cos^2(N\theta),&|\vec{\mathbf{b}}|^2\text{ is odd},\\
    \cos^2(N\theta),&|\vec{\mathbf{b}}|^2\text{ is even and }k\neq0.
\end{cases}
\end{aligned}
\end{equation}
Therefore, the QFI can be calculated according to Eq.~(10) of the main text that
\begin{equation}\label{eq:QFI for 2-LUI-RE GHZ}
\begin{aligned}
    &\mathcal F_{\mathrm{2-LUI-RE}}(\psi_{\mathrm{GHZ}})\\&=\frac{\big[\partial_{\theta}\cos^2(N\theta)\big]^2}{2^N}\left[\frac{1}{2^{N-1}+\cos^2(N\theta)}+\frac{2^{N-1}}{1-\cos^2(N\theta)}+\frac{2^{N-1}-1}{\cos^2(N\theta)}\right]=2N^2 \left[ 1 - \frac{\sin^2 (N\theta)}{\cos^2 (N\theta) + 2^{N-1}} \right].
\end{aligned}
\end{equation}
For $\theta\rightarrow0$, the GHZ state will give the result that
\begin{equation}\label{eq:HL limit}
    \lim_{\theta\rightarrow0}\mathcal{F}_{\mathrm{2-LUI-RE}}(\psi_{\mathrm{GHZ}})=\mathcal{F}_{\max}=2N^2,
\end{equation}
allowing the QFI to saturate the HL.

\section{Cases for Global Twirling }

In this section, we will discuss another case of global unitary invariant (GUI) states generated by taking the same global randomized rotation process to the two copies. We will give the general form of the 2-copy GUI state and then analyze the QFI for this type of state. Similar to the 2-LUI-RE protocol, we demonstrate that the 2-GUI-RE protocol proposed here, which just replaces the local randomized process in the 2-LUI-RE protocol by the global randomized process, also has the metrological power to overcome the decoherence-like noise caused by the RF misalignment. 

For the global twirling case, the GUI state is defined as $\tilde{\rho}_{\theta}=\Phi^{(2)}(\mathrm{P}_{\theta})=\int_{\text{Haar}}d\hat{U}\;\hat{U}^{\otimes2}\mathrm{P}_{\theta}\hat{U}^{\dagger\otimes2}$ with $\hat{U}$ a global unitary rotation acting on a single copy. It can be simplified according to the Schur-Weyl Duality as
\begin{equation}\label{eq:GUI}
    \tilde{\rho}_{\theta}=\frac{1}{d^{2N}-1}\Big(\hat{A}+\Tr(\hat{S}\mathrm{P}_{\theta})\hat{B}\Big),
\end{equation}
where $\hat{A}=\hat{I}-\hat{S}/d^N$ and $\hat{B}=\hat{S}-\hat{I}/d^N$ with $\hat{I}$ and $\hat{S}$ the global identity and SWAP operator respectively.

This GUI state is parameter-dependent on the term $\Tr(\hat{S}\mathrm{P}_{\theta})$ according to Eq.~(\ref{eq:GUI}), which is relevant to the encoding protocol. For the IE protocol proposed in \cite{Imai2024.arXiv.2410.10518}, we find that the term $\Tr(\hat{S}\mathrm{P}_{\theta})=\Tr(\rho_{\theta}^2)=1$ as $\rho_{\theta}$ is a pure state, which is independent on the parameter. Thus, the IE protocol here has no metrological power.

Now we focus on the 2-GUI-RE protocol, where we use two reversed encoding processes to the two copies instead of identical encoding process. And the trace term can be expressed as $\Tr(\hat{S}\mathrm{P}_{\theta})=\Tr(\rho_{\theta}\rho_{-\theta})$. To calculate the QFI of this GUI state, we first get the two different eigenvalues of this state that
\begin{equation}\label{eq:eigenvalue of GUI state}
\begin{aligned}
    \lambda_{1,2}&=\frac{1}{d^{2N}-1}\Big(1\mp\frac{1}{d^N}\Big)\big[1\pm\operatorname{Tr}(\rho_{\theta}\rho_{-\theta})\big],
\end{aligned}
\end{equation}
with the degeneracies of $d^N(d^N\pm1)/2$. Thus, the QFI can be calculated as
\begin{equation}\label{eq:QFI for 2-GUI-RE}
\begin{aligned}
    \mathcal F_{\mathrm{2-GUI-RE}}=\frac{d^N(d^N+1)}{2}\frac{(\partial_{\theta}\lambda_1)^2}{\lambda_1}+\frac{d^N(d^N-1)}{2}\frac{(\partial_{\theta}\lambda_2)^2}{\lambda_2}=\frac{\big[\partial_\theta \operatorname{Tr}(\rho_{\theta}\rho_{-\theta})\big]^2}{1-\big[\operatorname{Tr}(\rho_{\theta}\rho_{-\theta})\big]^2}.
\end{aligned}
\end{equation}

Then we prove that the QFI for the 2-GUI-RE protocol is also asymptotically optimal as $\theta\rightarrow0$. Here we define $|\psi_0\rangle$ as the initial pure state and $\rho_0=|\psi_0\rangle\langle\psi_0|$. After encoding, the states are $|\psi_\theta\rangle=e^{-i\hat{H}\theta}|\psi_0\rangle$ and $|\psi_{-\theta}\rangle=e^{i\hat{H}\theta}|\psi_0\rangle$. Then the QFI can be rewritten as
\begin{equation}\label{eq:pure state QFI for 2-GUI-RE}
    \mathcal F_{\mathrm{2-GUI-RE}}=\frac{\Big[\partial_{\theta}\big|\langle\psi_{\theta}|\psi_{-\theta}\rangle\big|^2\Big]^2}{1-\big|\langle\psi_{\theta}|\psi_{-\theta}\rangle\big|^4}.
\end{equation}
Further, we can use Taylor expansion for a sufficient small $\theta$ to expand the states to the second order of $\theta$ and get the result that
\begin{equation}\label{eq:inner product}
    \langle\psi_\theta|\psi_{-\theta}\rangle=1+2i\langle \hat{H}\rangle\theta-2\langle \hat{H}^2\rangle\theta^2+\mathcal O(\theta^3),
\end{equation}
where $\langle\cdot\rangle$ means $\langle\psi_0|\cdot|\psi_0\rangle$. Squaring this yields
\begin{equation}\label{eq:square inner product}
    \operatorname{Tr}(\rho_{\theta}\rho_{-\theta})=\big|\langle\psi_\theta|\psi_{-\theta}\rangle\big|^2=1-4\big(\langle \hat{H}^2\rangle-\langle \hat{H}\rangle^2\big)\theta^2+\mathcal O(\theta^3).
\end{equation}
Substituting into both the numerator and the denominator of Eq.~\eqref{eq:pure state QFI for 2-GUI-RE}, we can derive the limit value of the QFI as $\theta\rightarrow0$ that
\begin{equation}\label{eq:limit QFI for 2-GUI-RE}
\begin{aligned}
    \lim_{\theta\rightarrow0}\mathcal F_{\mathrm{2-GUI-RE}}=\frac{\Big[8\big(\langle \hat{H}^2\rangle-\langle \hat{H}\rangle^2\big)\theta\Big]^2}{8\big(\langle \hat{H}^2\rangle-\langle \hat{H}\rangle^2\big)\theta^2}=8\big(\langle \hat{H}^2\rangle-\langle \hat{H}\rangle^2\big)=\mathcal F_0.
\end{aligned}
\end{equation}
The expression of Eq.~(\ref{eq:limit QFI for 2-GUI-RE}) shows that the QFI can be asymptotically optimal for the 2-GUI-RE protocol when estimating a sufficiently small parameter $\theta$.

\begin{figure}[t]
    \centering
    \includegraphics[width=0.5\linewidth]{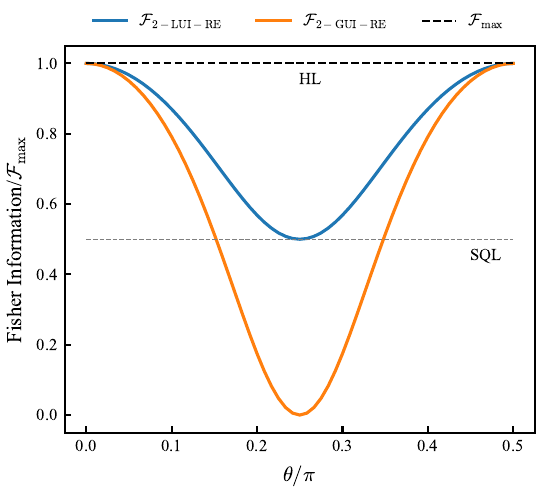}
    \caption{\textbf{Quantum Fisher Information for 2-LUI-RE and 2-GUI-RE Protocols.} The plot shows the QFIs of the two protocols as a function of the encoded parameter $\theta$, normalized by the theoretical maximum $\mathcal{F}_{\max}=2N^2$ which means the Heisenberg limit (HL, black dashed line). The simulation uses the initial GHZ state and the encoding process as $\hat{H}=\frac{1}{2}\sum_{i}Z_i$ with $N=2$. The blue solid line shows the QFI for the 2-LUI-RE protocol, $\mathcal{F}_{\text{2-LUI-RE}}$, which remains above the stand quantum limit (SQL, grey dashed line) and saturate the HL at optimal points. The orange solid line shows the QFI for the 2-GUI-RE protocol, $\mathcal{F}_{\text{2-GUI-RE}}$. Compared with the $\mathcal{F}_{\text{2-LUI-RE}}$, the $\mathcal{F}_{\text{2-GUI-RE}}$ can be lower than the SQL at some points, even though it can also saturate the HL at the optimal points.}
    \label{fig:QFI}
\end{figure}

In this paragraph, we take Example 3 of the main text here to demonstrate these results more clearly and then compare them with the 2-LUI-RE protocol. We also prepare two copies of the initial GHZ state $|\psi_{\mathrm{GHZ}}\rangle=\frac{1}{\sqrt{2}}(|H_1H_2\cdots H_N\rangle+|V_1V_2\cdots V_N\rangle)$ and consider the encoding Hamiltonian $\hat{H}=\frac{1}{2}\sum_{i}Z_i$. In this case, we have $\operatorname{Tr}(\rho_{\theta}\rho_{-\theta})=\cos^2(N\theta)$. Therefore, the QFI can be derived according to Eq.~(\ref{eq:QFI for 2-GUI-RE}) that
\begin{equation}\label{eq:QFI for 2-GUI-RE GHZ}
    \mathcal F_{\mathrm{2-GUI-RE}}(\psi_{\mathrm{GHZ}})=\frac{\big[\partial_{\theta}\cos^2(N\theta)\big]^2}{1-\cos^4(N\theta)}=\frac{4N^2\cos^2(N\theta)}{1+\cos^2(N\theta)}.
\end{equation}
For $\theta\rightarrow0$, we will get the same result with the 2-LUI-RE protocol that
\begin{equation}\label{eq:HL limit GUI}
    \lim_{\theta\rightarrow0}\mathcal{F}_{\mathrm{2-GUI-RE}}(\psi_{\mathrm{GHZ}})=\mathcal{F}_{\max}=2N^2,
\end{equation}
which demonstrate that the 2-GUI-RE protocol is also asymptotically optimal, whose QFI can saturate to the HL for initial GHZ states. Moreover, we compare the 2-GUI-RE and 2-LUI-RE protocols for GHZ states in Fig.~\ref{fig:QFI}. Compared with the 2-LUI-RE protocol, the 2-GUI-RE protocol has mainly two disadvantages. First, according to Fig.~\ref{fig:QFI}, although the QFI can saturate the HL for a sufficiently small $\theta$, it can be worse than the SQL and even decrease to zero at some $\theta$. In contrast, the QFI for the 2-LUI-RE protocol is always above the SQL, which means a better precision than any classical protocols. Second, since the randomized unitary process is global, it can hardly be used as a distributed metrology protocol.

\section{Measurement}

From the perspective of estimation theory, the classical Fisher information (CFI) derived from the measurement results limits the estimation precision of the parameter $\theta$ through the Cram$\acute{\text{e}}$r-Rao bound (CRB). If we use a given positive operator-valued measure (POVM) $\{\hat{\Pi}_x\}$ to measure the LUI state $\tilde{\rho}_{\theta}$, we can get the probability $p(x|\theta)=\operatorname{Tr}(\hat{\Pi}_x\tilde{\rho}_{\theta})$. Then the CFI can be calculated as
\begin{equation}\label{eq:CFI formular}
    F=\int dx\;p(x|\theta)\big[\partial_{\theta}\ln p(x|\theta)\big]^2=\int dx\;\frac{\big[\partial_{\theta}p(x|\theta)\big]^2}{p(x|\theta)}.
\end{equation}

In the following parts, we will take several common measurements and calculate the CFI by Eq.~\eqref{eq:CFI formular} for the reversed encoding protocol.

\subsection{Local Randomized Measurement on the Computational Basis}

First, we consider the case of local randomized measurement (LRM) on computational bases. We define $\hat{\Pi}_{s_\text{A},s_\text{B}}=|s_\text{A}\rangle\langle s_\text{A}|\otimes|s_\text{B}\rangle\langle s_\text{B}|$ as the measurement operators with $|s_\text{A}\rangle$ and $|s_\text{B}\rangle$ the $d$-dimension computational bases of copy A and B respectively. The total number of computational bases is $d^{2N}$ and we have $\sum_{s_\text{A},s_\text{B}}\hat{\Pi}_{s_\text{A},s_\text{B}}=\hat{I}$. We need to do the local randomized process to $\hat{\Pi}_{s_\text{A},s_\text{B}}$. As the randomized process acting on the measurement operator is equivalent to that acting on the state, this randomized measurement process is equivalent to the direct measurement (DM) on computational bases to the LUI state under local randomized process. For convenience, we rewrite the measurement operator in the product form that $\hat{\Pi}_{s_\text{A},s_\text{B}}=\bigotimes_{i=1}^N|s_\text{A}s_\text{B}\rangle_i\langle s_\text{A}s_\text{B}|$, where $|s_\text{A}s_\text{B}\rangle_i$ is the computational basis of the $i$-th particle for both copies. Then the probabilities can be derived as
\begin{equation}\label{eq:probability for LRM}
    p(\theta)=\operatorname{Tr}(\hat{\Pi}_{s_\text{A},s_\text{B}}\tilde{\rho}_{\theta})=\Big(\frac{1}{d^2-1}\Big)^N\sum_{\vec{\mathbf{a}}}\operatorname{Tr}(\rho_{\theta,\vec{\mathbf{a}}}\rho_{-\theta,\vec{\mathbf{a}}})\prod_{i:a_i=0}\Big(1-\frac{1}{d}\big|\langle s_\text{A}|s_\text{B}\rangle_i\big|^2\Big)\prod_{j:a_j=1}\Big(\big|\langle s_\text{A}|s_\text{B}\rangle_j\big|^2-\frac{1}{d}\Big).
\end{equation}
Therefore, the CFI can be calculated as
\begin{equation}\label{eq:CFI for LRM}
    F^{\mathrm{LRM}}=\sum_{s_\text{A},s_\text{B}}\frac{\big[\partial_\theta p(\theta)\big]^2}{p(\theta)}=\frac{1}{(d+1)^N}\sum_{n=0}^N\binom{N}{n}\frac{\big[\partial_{\theta}\operatorname{Tr}(\rho_{\theta}\rho_{-\theta})\big]^2}{d^n+(-1)^n\operatorname{Tr}(\rho_{\theta}\rho_{-\theta})},
\end{equation}
for both 2-GUI-RE and 2-LUI-RE. Also, we consider the example of GHZ state, where the CFI can be derived as
\begin{equation}\label{eq:CFI for LRM GHZ}
    F^{\mathrm{LRM}}=\frac{4N^2}{(d+1)^N}\sum_{n=0}^N\binom{N}{n}\frac{\cos^2(N\theta)\sin^2(N\theta)}{d^n+(-1)^n\cos^2(N\theta)}.
\end{equation}
Therefore, LRM is not optimal as well according to the result. For the case of $d=2$ and $N=2$, the maximum CFI is $F^{\mathrm{LRM}}_{\max}=0.990=12.4\%\mathcal F_{\max}$. For a general situation, we take the derivative of Eq.~\eqref{eq:CFI for LRM GHZ} and get the following condition to maximize the CFI that
\begin{equation}\label{eq:sum zero}
    \sum_{n=0}^N\binom{N}{n}\frac{(-1)^{n+1}\cos^4(N\theta)-2d^n\cos^2(N\theta)+d^n}{(-1)^n\cos^4(N\theta)+d^n}=0.
\end{equation}
If $N$ is large enough, we can simplify Eq.~\eqref{eq:sum zero} and get the condition $\cos^2(N\theta)\approx1/2$, with which the maximum CFI has the approximate result that
\begin{equation}\label{eq:large N max CFI for LRM}
    F^{\mathrm{LRM}}_{\max}\approx\frac{N^2}{(d+1)^N}\sum_{n=0}^N\binom{N}{n}\frac{1}{d^n}=\frac{1}{2d^N}\mathcal F_{\max}=\frac{N^2}{d^N}\;\;\;\;(\text{large}\;N),
\end{equation}
which means LRM cannot preserve the HL scaling.

\begin{figure}[t]
    \centering
    \includegraphics[width=0.5\linewidth]{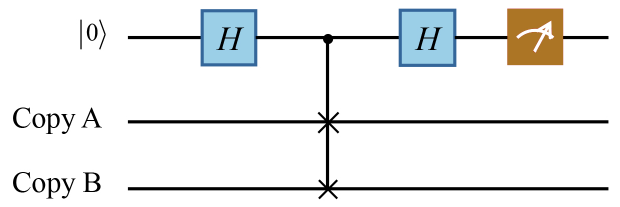}
    \caption{\textbf{The General Process of SWAP Test.} We prepare an ancillary qubit in $|0\rangle$ and apply a Hadamard gate to it. Then we perform a controlled-SWAP operation between copies A and B with this ancilla. After that, another Hadamard gate is applied to the ancilla, and a measurement is taken to get $p_{\pm}$.}
    \label{fig:SWAP}
\end{figure} 

\subsection{Global Randomized Measurement on the Computational Basis}

Here we consider the case of global randomized measurement (GRM) on computational bases, where we need to do the global randomized process to $\hat{\Pi}_{s_\text{A},s_\text{B}}$ that
\begin{equation}\label{eq:Schur-Weyl for global Z measurement}
\begin{aligned}
    \Phi^{(2)}(\hat{\Pi}_{s_\text{A},s_\text{B}})&=\frac{1}{d^{2N}-1}\left\{\Big[1-\frac{1}{d^N}\operatorname{Tr}(\hat{S}\hat{\Pi}_{s_\text{A},s_\text{B}})\Big]\hat{I}+\Big[\operatorname{Tr}(\hat{S}\hat{\Pi}_{s_\text{A},s_\text{B}})-\frac{1}{d^N}\Big]\hat{S}\right\}\\&=\begin{cases}\frac{1}{d^N(d^N+1)}\big(\hat{I}+\hat{S}\big),&|s_\text{A}\rangle=|s_\text{B}\rangle,\\\frac{1}{d^{2N}-1}\big(\hat{I}-\frac{1}{d^N}\hat{S}\big),&|s_\text{A}\rangle\neq|s_\text{B}\rangle.\end{cases}
\end{aligned}
\end{equation}
Under this measurement, the probabilities can be derived using $p(\theta)=\operatorname{Tr}\big[\Phi^{(2)}(\hat{\Pi}_{s_\text{A},s_\text{B}})\tilde{\rho}_{\theta}\big]$ as
\begin{equation}\label{eq:probabilities}
\begin{aligned}
    &p_1(\theta)=\frac{1}{d^N(d^N+1)}\operatorname{Tr}\big[(\hat{I}+\hat{S})\tilde{\rho}_{\theta}\big]=\frac{1}{d^N(d^N+1)}\big[1+\operatorname{Tr}(\rho_{\theta}\rho_{-\theta})\big],&&|s_\text{A}\rangle=|s_\text{B}\rangle,\\&p_2(\theta)=\frac{1}{d^{2N}-1}\operatorname{Tr}\Big[\big(\hat{I}-\frac{1}{d^N}\hat{S}\big)\tilde{\rho}_{\theta}\Big]=\frac{1}{d^{2N}-1}\Big[1-
    \frac{1}{d^N}\operatorname{Tr}(\rho_{\theta}\rho_{-\theta})\Big],&&|s_\text{A}\rangle\neq|s_\text{B}\rangle,
\end{aligned}
\end{equation}
with the degeneracy of $d^N$ and $d^N(d^N-1)$ respectively. Therefore, the CFI can be derived as
\begin{equation}\label{eq:CFI for GRM}
    F^{\mathrm{GRM}}=d^N\frac{\big[\partial_{\theta}p_1(\theta)\big]^2}{p_1(\theta)}+d^N(d^N-1)\frac{\big[\partial_{\theta}p_2(\theta)\big]^2}{p_2(\theta)}=\frac{\big[\partial_{\theta}\operatorname{Tr}(\rho_{\theta}\rho_{-\theta})\big]^2}{d^N+(d^N-1)\operatorname{Tr}(\rho_{\theta}\rho_{-\theta})-\big[\operatorname{Tr}(\rho_{\theta}\rho_{-\theta})\big]^2},
\end{equation}
for both 2-GUI-RE and 2-LUI-RE . To demonstrate this result intuitively, we consider the example of the GHZ state, where the CFI is in the form of
\begin{equation}\label{eq:CFI for GRM GHZ}
    F^{\mathrm{GRM}}=\frac{4N^2\cos^2(N\theta)\sin^2(N\theta)}{d^N+(d^N-1)\cos^2(N\theta)-\cos^4(N\theta)}.
\end{equation}
This result is always less than the QFI for GHZ states as a $d^N$ term appears in the denominator. Therefore, GRM is not optimal. For the simple case of $d=2$ and $N=2$ (two qubits for each copy), we can obtain the maximum CFI as $F^{\mathrm{GRM}}_{\max}=0.769=9.6\%\mathcal F_{\max}$. For a general situation, the maximum CFI has the form of 
\begin{equation}\label{eq:max CFI for GRM GHZ}
    F^{\mathrm{GRM}}_{\max}=4N^2\frac{(3d^N-2)\cos^2(N\theta)-d^N}{(d^{2N}-d^N+2)\cos^2(N\theta)+d^{2N}-3d^N},
\end{equation}
with $\cos^2(N\theta)=\frac{\sqrt{2d^N(d^N-1)}-d^N}{d^N-2}$ to get the maximum CFI. If $N$ is large enough, we have $\cos^2(N\theta)\approx\sqrt{2}-1$ to obtain the maximum CFI approximately that
\begin{equation}\label{eq:large N max CFI for GRM}
    F^{\mathrm{GRM}}_{\max}\approx\frac{6-4\sqrt{2}}{d^N}\mathcal F_{\max}=(12-8\sqrt{2})\frac{N^2}{d^N}\;\;\;\; (\text{large}\; N),
\end{equation}
which means GRM cannot preserve the HL scaling as well.

\subsection{SWAP Test}

Now we consider the case of SWAP test \cite{Buhrman2001.PhysRevLett.87.167902}, the process of which is shown in Fig.~\ref{fig:SWAP}. 
If we take a global SWAP test (GST) where we use a single ancillary qubit to control the swap operation across all sites, measuring the ancillary qubit will give the two measurement results as $p_{\pm}(\theta)=\frac{1}{2}\big(1\pm\langle\hat{S}\rangle\big)$ where $\langle\hat{S}\rangle=\operatorname{Tr}(\hat{S}\tilde{\rho}_{\theta})=\Tr(\hat{S}\mathrm{P}_{\theta})$. According to Eq.~(5) of the main text, the expected value of $\hat{S}$ can be derived as
\begin{equation}\label{eq:S expect}
    \langle\hat{S}\rangle=\frac{1}{2^N}\sum_{\vec{\mathbf{a}}}\operatorname{Tr}(\rho_{\theta,\vec{\mathbf{a}}}\rho_{-\theta,\vec{\mathbf{a}}})\sum_{\vec{\mathbf{b}}}(-1)^{\vec{\mathbf{a}}\cdot\vec{\mathbf{b}}+|\vec{\mathbf{b}}|^2}=\operatorname{Tr}(\rho_{\theta}\rho_{-\theta}),
\end{equation}
which is the same with the case of 2-GUI-RE. Therefore, the CFI can be derived as
\begin{equation}\label{eq:CFI for ST}
    F^{\mathrm{GST}}=\frac{\big[\partial_{\theta}p_0(\theta)\big]^2}{p_0(\theta)}+\frac{\big[\partial_{\theta}p_1(\theta)\big]^2}{p_1(\theta)}=\frac{\big[\partial_{\theta}\operatorname{Tr}(\rho_{\theta}\rho_{-\theta})\big]^2}{1-\big[\operatorname{Tr}(\rho_{\theta}\rho_{-\theta})\big]^2}.
\end{equation}
Then we can easily find that this CFI is equal to $\mathcal F_{\mathrm{2-GUI-RE}}$ , and we still have the limit that $\lim_{\theta\rightarrow0}F^{\mathrm{GST}}=\mathcal F_0$, which means that the GST is asymptotically optimal for 2-LUI-RE protocol.

If we take the local SWAP test (LST) shown in Fig.~4(b) of the main text, the information contained in $\langle\hat{S}_{\vec{\mathbf{a}}}\rangle=\Tr(\hat{S}_{\vec{\mathbf{a}}}\tilde{\rho}_{\theta})=\Tr(\hat{S}_{\vec{\mathbf{a}}}\mathrm{P}_{\theta})$ can be extracted by measuring all the ancillary qubits and then getting the probabilities $p_{\vec{\mathbf{b}}}$ with $\vec{\mathbf{b}}$ an $N$-bit binary string representing the measurement basis on each ancillary qubit. As shown in Fig.~4(b) of the main text, after the second Hadamard gate for all the ancillary qubits, the state of the whole system and ancillary qubits can be expressed as
\begin{equation}\label{eq:whole state}
\begin{aligned}
    |\Phi_{\theta}\rangle&=\Big\{\bigotimes_{i=1}^N\frac{1}{2}\big[(|0\rangle_i+|1\rangle_i)\otimes\hat{I}_i+(|0\rangle_i-|1\rangle_i)\otimes\hat{S}_i\big]\Big\}|\Psi_{\theta}\rangle\\&=\frac{1}{2^N}\sum_{\vec{\mathbf{b}}}|\vec{\mathbf{b}}\rangle\otimes\prod_{i:b_i=0}(\hat{I}_i+\hat{S}_i)\prod_{j:b_j=1}(\hat{I}_i-\hat{S}_i)|\Psi_{\theta}\rangle,
\end{aligned}
\end{equation}
where $|0\rangle_i$ and $|1\rangle_i$ is the two measurement bases on the $i$-th ancillary qubit and we define $\mathrm{P}_{\theta}=|\Psi_{\theta}\rangle\langle\Psi_{\theta}|$. Then the probabilities can be derived as
\begin{equation}\label{eq:probability for LST}
    p_{\vec{\mathbf{b}}}(\theta)=\frac{1}{2^N}\langle\prod_{i:b_i=0}(\hat{I}+\hat{S}_i)\prod_{j:b_j=1}(\hat{I}-\hat{S}_j)\rangle=\frac{1}{2^N}\sum_{\vec{\mathbf{a}}}(-1)^{\vec{\mathbf{a}}\cdot\vec{\mathbf{b}}}\langle\hat{S}_{\vec{\mathbf{a}}}\rangle.
\end{equation}
Therefore, the CFI for LST can be derived as
\begin{equation}
    F^{\text{LST}}=\sum_{\vec{\mathbf{b}}}\frac{\big[\partial_{\theta}p_{\vec{\mathbf{b}}}(\theta)\big]^2}{p_{\vec{\mathbf{b}}}(\theta)}=\frac{1}{2^N}\sum_{\vec{\mathbf{b}}}\frac{\Big[\sum_{\vec{\mathbf{a}}}(-1)^{\vec{\mathbf{a}}\cdot\vec{\mathbf{b}}}\partial_{\theta}\operatorname{Tr}(\rho_{\theta,\vec{\mathbf{a}}}\rho_{-\theta,\vec{\mathbf{a}}})\Big]^2}{\sum_{\vec{\mathbf{a}}}(-1)^{\vec{\mathbf{a}}\cdot\vec{\mathbf{b}}}\operatorname{Tr}(\rho_{\theta,\vec{\mathbf{a}}}\rho_{-\theta,\vec{\mathbf{a}}})}=\mathcal{F}_{\text{2-LUI-RE}},
\end{equation}
which means the LST is the optimal strategy to get all the QFI for any $\theta$.

\subsection{Local Bell-state Measurement for Qubit System}

We have proved that LST is the optimal measurement for 2-LUI-RE protocol. Here we will prove that the local Bell-state measurement (LBM) is also optimal for qubit systems ($d=2$). First we can express the four local Bell states in the reference frame of the $i$-th site that
\begin{equation}\label{eq:Bell sates}
\begin{aligned}
    |\phi_+\rangle_i&=\frac{1}{\sqrt{2}}\big(|H_iH_i\rangle+|V_iV_i\rangle\big),\\|\phi_-\rangle_i&=\frac{1}{\sqrt{2}}\big(|H_iH_i\rangle-|V_iV_i\rangle\big),\\|\psi_+\rangle_i&=\frac{1}{\sqrt{2}}\big(|H_iV_i\rangle+|V_iH_i\rangle\big),\\|\psi_-\rangle_i&=\frac{1}{\sqrt{2}}\big(|H_iV_i\rangle-|V_iH_i\rangle\big).
\end{aligned}
\end{equation}
We can easily find that these four states are just the eigenstates of the local SWAP operator $\hat{S}_i$ with the first three symmetric states and the last one anti-symmetric states. Then the eigenstates of the global SWAP operator $\hat{S}$ can be expressed as $|\varphi_{\hat{S}}\rangle:=\bigotimes_{i=1}^{N}|\varphi_{\hat{S}_i}\rangle$ where $|\varphi_{\hat{S}_i}\rangle$ can take any of the four Bell states. The measurement operators can be defined as $\hat{\Pi}_{\varphi_{\hat{S}}}:=|\varphi_{\hat{S}}\rangle\langle\varphi_{\hat{S}}|=\bigotimes_{i=1}^N |\varphi_{\hat{S}_i}\rangle\langle\varphi_{\hat{S}_i}|$ with $\sum_{\varphi_{\hat{S}}}\hat{\Pi}_{\varphi_{\hat{S}}}=\hat{I}$. And we can express the measurement probabilities for the 2-LUI-RE protocol that
\begin{equation}\label{eq:probability for Bell measurement LUI}
\begin{aligned}
    p(\theta)=\operatorname{Tr}(\hat{\Pi}_{\varphi_{\hat{S}}}\tilde{\rho}_{\theta})&=\frac{1}{3^N}\sum_{\vec{\mathbf{a}}}\operatorname{Tr}(\rho_{\theta,\vec{\mathbf{a}}}\rho_{-\theta,\vec{\mathbf{a}}})\prod_{i:a_i=0}\operatorname{Tr}(\hat{\Pi}_{\varphi_{\hat{S}_i}}\hat{A}_i)\prod_{j:a_j=1}\operatorname{Tr}(\hat{\Pi}_{\varphi_{\hat{S}_j}}\hat{B}_j)\\&=\frac{1}{3^N}\Big(\frac{1}{2}\Big)^{N-|\vec{\mathbf{b}}|^2}\Big(\frac{3}{2}\Big)^{|\vec{\mathbf{b}}|^2}\sum_{\vec{\mathbf{a}}}(-1)^{\vec{\mathbf{a}}\cdot\vec{\mathbf{b}}}\operatorname{Tr}(\rho_{\theta,\vec{\mathbf{a}}}\rho_{-\theta,\vec{\mathbf{a}}}),
\end{aligned}
\end{equation}
with the degeneracy of $3^{N-|\vec{\mathbf{b}}|^2}$. Then the CFI is
\begin{equation}\label{eq:CFI for MESO 2-LUI-RE}
    F^{\text{LBM}}=\sum_{\vec{\mathbf{b}}}3^{N-|\vec{\mathbf{b}}|^2}\frac{\big[\partial_{\theta}p(\theta)\big]^2}{p(\theta)}=\frac{1}{2^N}\sum_{\vec{\mathbf{b}}}\frac{\Big[\sum_{\vec{\mathbf{a}}}(-1)^{\vec{\mathbf{a}}\cdot\vec{\mathbf{b}}}\partial_{\theta}\operatorname{Tr}(\rho_{\theta,\vec{\mathbf{a}}}\rho_{-\theta,\vec{\mathbf{a}}})\Big]^2}{\sum_{\vec{\mathbf{a}}}(-1)^{\vec{\mathbf{a}}\cdot\vec{\mathbf{b}}}\operatorname{Tr}(\rho_{\theta,\vec{\mathbf{a}}}\rho_{-\theta,\vec{\mathbf{a}}})}.
\end{equation}
And we have $F^{\text{LBM}}=\mathcal{F}_{\text{2-LUI-RE}}$, which means LBM is the optimal measurement for the 2-LUI-RE protocol in qubit systems.

This result provides us with a valid protocol for reference-frame (RF) independent distributed sensing in qubit systems shown in Fig.~4(c) of the main text. Local randomized unitary rotations overcome the decoherence effect caused by the absence of a shared RF, and two copies ensure that the LUI state contains the information of the parameter. The reversed encoding process circumvents the no-go theorem to get a non-zero QFI only with local operations, while LBM can obtain the whole QFI of the 2-copy LUI state to achieve the HL scaling with GHZ states.

\bibliography{references}